\documentclass[reprint,twoside]{revtex4-1} %
\usepackage{hyperref} %

\usepackage{amsmath}	
\usepackage[dvipdf,]{graphicx}
\usepackage{color} %
\usepackage{bm} %
\usepackage{subcaption} %
\usepackage{tikz}	%
\usetikzlibrary{arrows,shapes,positioning} %
\usetikzlibrary{decorations.markings} %
\tikzstyle arrowstyle=[scale=1]	 %
\usetikzlibrary{quotes,angles}	%
\usepackage{pgfplots}
\usepackage{caption}  %
\newcommand{\dc}[1]{\textcolor{red}{[Comment: \textsc{#1}]}} %
\newcommand{\bold}[1]{\textbf{#1}} %
\newcommand{\ttm}[1]{\textrm{#1}} %

\newcommand{\aloop}{$\langle a \rangle$ loop}
\newcommand{\aloops}{$\langle a \rangle$ loops}
\newcommand{\cloop}{$\langle c \rangle$ loop}
\newcommand{\cloops}{$\langle c \rangle$ loops}
\newcommand{\cploop}{$\langle c/2+ p \rangle$ loop}
\newcommand{\cploops}{$\langle c/2+p \rangle$ loops}
\newcommand{\chloop}{$\langle c/2 \rangle$ loop}
\newcommand{\chloops}{$\langle c/2 \rangle$ loops}

\newlength{\figwidth}
\newlength{\widefigwidth}
\begin{document}
\title{%
An Atomistic Modelling Study of the Properties of Dislocation Loops in Zirconium\\
}
\vspace{2cm}
\author{R.Hulse} %
\email[Contact: ]{rory.hulse@manchester.ac.uk}
\author{C.P.Race.} %
\affiliation{Department of Materials, University of Manchester, Manchester, M13 9PL, United Kingdom}

\maketitle
\thispagestyle{empty}

\section{Abstract}

\noindent  Neutron irradiation progressively changes the properties of zirconium alloys: they harden and their average c/a lattice parameter ratio decreases with fluence \cite{was2017fundamentals,onimus2012radiation,buckley1962properties,carpenter1981irradiation}.  The bombardment by neutrons produces point defects, which evolve into dislocation loops that contribute to a non-uniform growth phenomenon called irradiation-induced growth (IIG).  To gain insights into these dislocation loops in Zr we studied them using atomistic simulation.  We constructed and relaxed dislocation loops of various types.  We find that the energies of \aloops{} on different habit planes are similar, but our results indicate that they are most likely to form on the 1st prismatic plane and then reduce their energy by rotating onto the 2nd prismatic plane.  By simulating loops of different aspect ratios, we find that, based on energetics alone, the shape of \aloops{} does not depend on character, and that these loops become increasingly elliptical as their size increases.  Additionally, we find that interstitial \cploops{} and vacancy \cloops{} are both energetically feasible and so the possibility of these should be considered in future work.  Our findings offer important insights into loop formation and evolution, which are difficult to probe experimentally.

\section{Introduction}	\label{sec:intro}	%

\noindent Zirconium alloys are used in light water nuclear power reactors in several applications, in particular for nuclear fuel cladding.  These reactors operate at temperatures from around 350~K to about 580~K \cite{griffiths1987formation}.  At these temperatures $\alpha$-Zr has a hexagonal close-packed (HCP) crystal structure \cite{liu2009experimental}.  Recrystallised Zr alloy guide tubes and grids suffer from irradiation-induced growth (IIG), which occurs in three phases: initial,  steady and breakaway \cite{holt1986c}.  
IIG begins with the initial growth phase, where there is rapid growth up to a fluence of around $10^{25}$~n/m$^2$.  After this, the growth curve gradient flattens and there is a long phase where little or no growth occurs.  This is known as the steady growth phase.  Finally, at around $3 \times 10^{25}$~n/m$^2$, the growth rate becomes rapid and this is the breakaway growth phase \cite{holt1986c,carpenter1988irradiation}.  Breakaway growth is the most detrimental as in this regime the cladding rapidly elongates.  IIG constrains the design of fuel assemblies, so an IIG resistant alloy would allow for more freedom in their design.  IIG is an anisotropic growth and is stochastic.  This can cause the fuel rods to buckle, leading to problems such as hot spots \cite{pickman1975interactions}.  Note that IIG is one phenomenon that may contribute to fuel rod deformation, but other phenomena, such as pellet cladding interaction, may also contribute \cite{rossiter2012understanding}.

\begin{figure}[htbp!]\begin{center}
{\includegraphics[width=\figwidth]{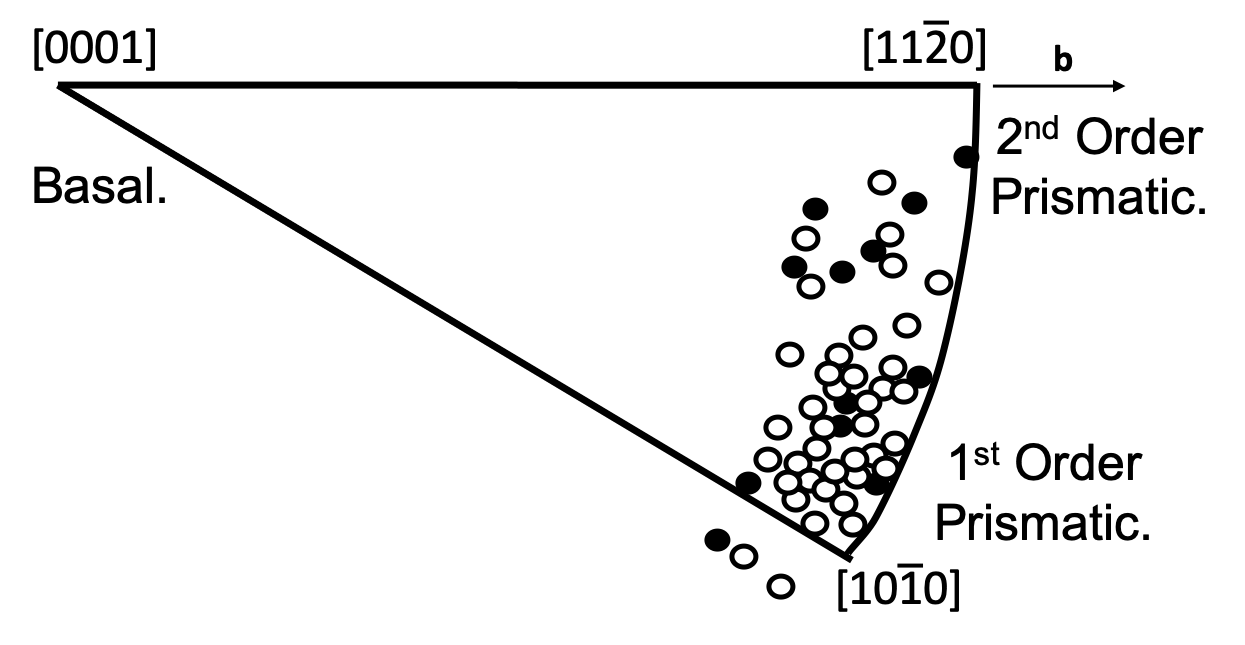}}
\newline
\caption{The \aloop{} habit planes in a sample of neutron-irradiated zone-refined Zr.  Empty and filled circles represent vacancy loops and interstitial loops respectively.  (Minimally adapted from Jostsons et al.~\cite{jostsons1977nature}, with permission from Elsevier.)}
\label{fig:aLoopPlanes}
\end{center}
\end{figure}

\begin{figure}[!ht]\begin{center}
{\includegraphics[width=\figwidth]{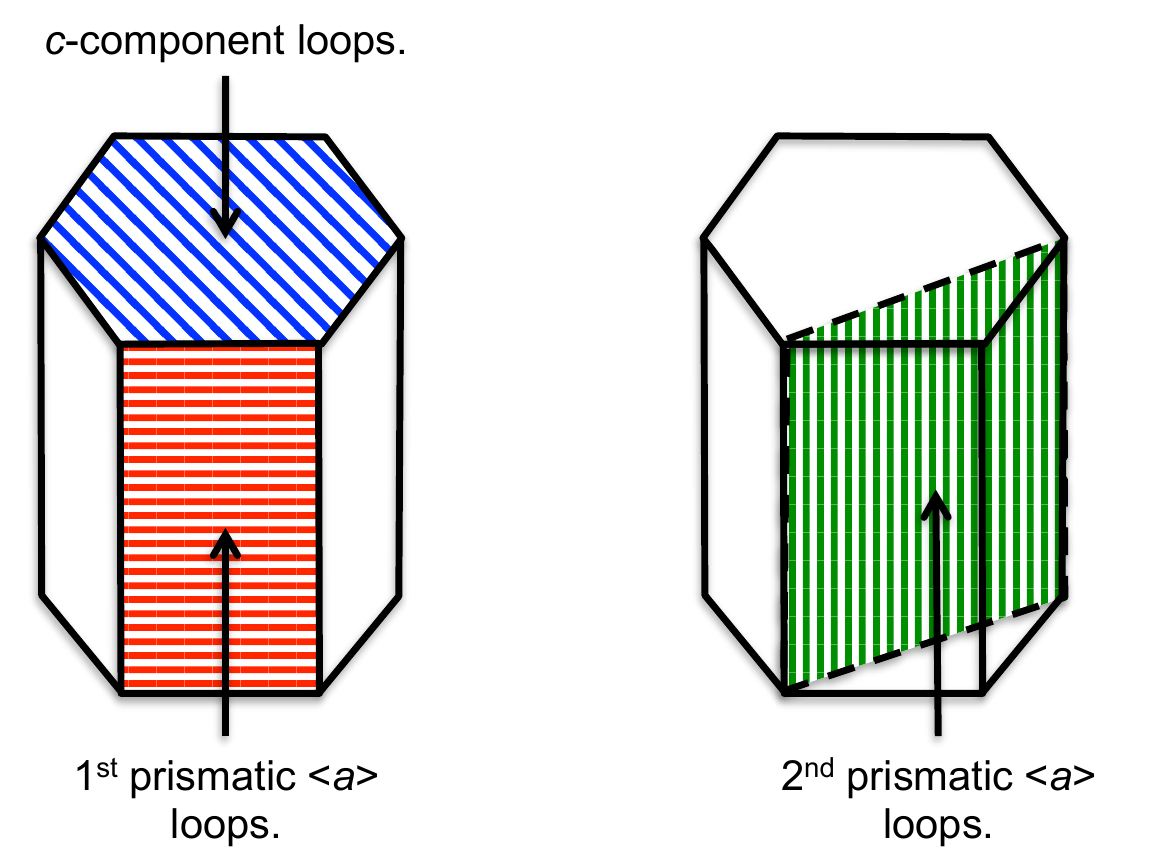}}
\caption{A schematic illustration of the orientation of key crystallographic planes with respect to the hexagonal symmetry of $\alpha$-Zr.} %
\label{fig:unitCell}
\end{center}
\end{figure}

The design of growth resistant alloys would be greatly aided by a comprehensive understanding of the microscopic mechanisms behind IIG.
A promising explanation of IIG provided by Buckley postulates that beyond its initial stage, IIG is primarily caused by irradiation-induced dislocation loops \cite{buckley1962properties}.  In the early stages of irradiation, dislocation loops with burgers vector $1/3 \langle1 1 \bar{2} 0\rangle$ are seen \cite{holt1986c,kelly1973characterization}.  These are termed \aloops{} and are of either vacancy or interstitial character \cite{kelly1973characterization}.  In Zr irradiated (E~$>$~1~MeV) at 668~K, Jostsons et al.~observed that 66$\%$ of the \aloops{} were vacancy in character at a fluence of 6.4~$\times$~10$^{19}$~n~cm$^{-2}$~\cite{jostsons1977nature}.  At a higher fluence of 1.8~$\times$~10$^{20}$~n~cm$^{-2}$, they characterised approximately equal numbers of vacancy and interstitial \aloops{}~\cite{jostsons1977nature}.  Northwood et al.~irradiated single crystal Zr to 1.8~$\times$~10$^{20}$~n~cm$^{-2}$ (E~$>$~1~MeV) at $\sim$500~K and characterised 44$\%$ of the \aloops{} as vacancy type~\cite{northwood1976dislocation}.  \aloops{} inhabit a variety of planes, clustered around the 1st prismatic planes \cite{jostsons1977nature,kelly1973characterization}, as shown in Fig.~\ref{fig:aLoopPlanes}.  At higher doses, irradiation-induced dislocation loops with a burgers vector containing a component in the $c$ direction, [0001], appear.  Holt and Gilbert believe that these $c$-component loops cause breakaway IIG \cite{holt1986c}.  $c$-component loops seen in irradiated Zr have been determined by transmission electron microscopy (TEM) to have a burgers vector of $1/6 \langle2 \bar{2} 0 3\rangle$ and are vacancy in character \cite{griffiths1987formation}.  For the remainder of this paper these $c$-component loops will be referred to as \cploops{} and they inhabit the basal planes \cite{griffiths1987formation}.  Also of interest are $c$-component loops with burgers vector $\langle 0 0 0 {2}\rangle$, which are termed \chloops{} and those with burgers vector $\langle0 0 0 1\rangle$ that are termed \cloops{}.  The reason \cloops{} are of interest is that they were observed in a TEM study by Griffiths \cite{griffiths1987formation}.  \chloops{} were observed in an earlier study by Griffiths et al.~\cite{griffiths1984anisotropic} and may be the precursors to \cploops \cite{hull2001introduction}.  Note that the $p$ in $\langle c/2+ p \rangle$ denotes the partial dislocation that transforms \chloops{} to \cploops{}.  The planes inhabited by the loops most pertinent to this study are shown schematically in Fig.~\ref{fig:unitCell}.

\subsection{Irradiation Induced Growth}
\noindent Jostsons et al.~observed that irradiated Zr single crystals elongated in the $a$-directions and shrank in the $c$-directions \cite{jostsons1977nature}.  Mechanical processing of Zr alloy fuel cladding induces a strong texture \cite{baron1990interlaboratories,tucker1984high}, where the majority of the basal plane normals are oriented in the radial direction for Zircaloy-2 fuel tube \cite{fidleris1987overview}.  An increase in the lengths of crystal grains in the $a$-direction, combined with their strong orientational order, causes macroscopic growth.

Carpenter et al.~reported behaviour consistent with Buckley's postulated IIG explanation that \aloops{} of interstitial character are a cause of $a$-direction expansion and \cploops{} are a cause of $c$-direction shrinkage \cite{carpenter1981irradiation,{buckley1962properties}}.\\

\subsection{Dislocation Loop Shape}
\begin{figure}[!ht]\begin{center}
{\includegraphics[width=\figwidth]{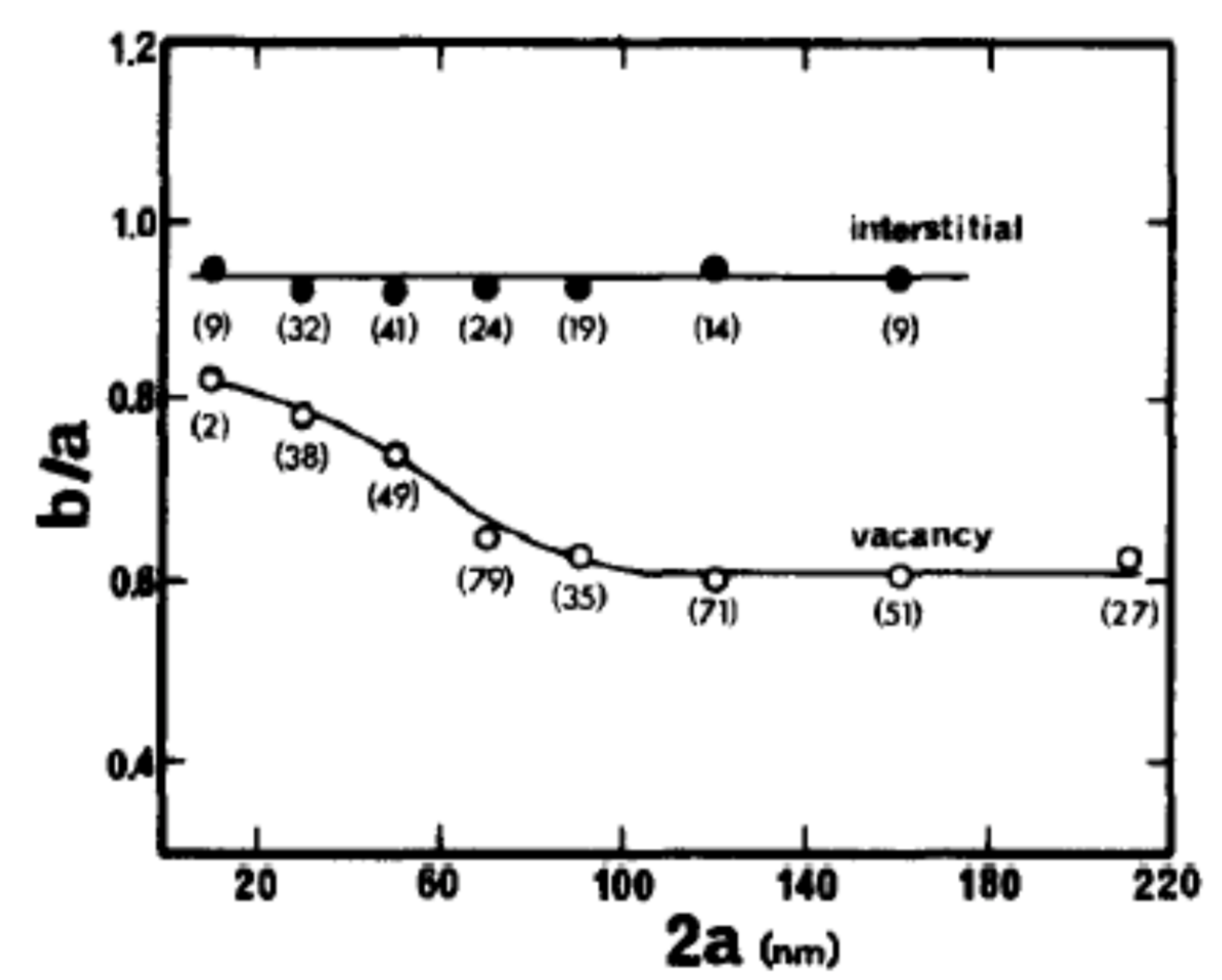}}	%
\caption{The ellipticities of \aloops{} as a function of loop size, where the numbers in parentheses are the quantity of loops observed.  In this plot, `2a' is the ellipse major axis and `2b' is the ellipse minor axis.  (Reproduced from Jostsons et al.~\cite{jostsons1977nature}, with permission from Elsevier.).}
\label{fig:ellipLoopsPlot}
\end{center}
\end{figure}

\noindent TEM images of \aloops{} in irradiated Zr alloy samples, show that these loops have an elliptical shape \cite{jostsons1977nature,northwood1979characterization}.  The major axis of elliptical \aloops{} lies parallel to the $c$-direction and the minor axis lies in the basal plane \cite{jostsons1977nature}.  Ellipticities of various \aloops{}, of vacancy and interstitial type, are shown in Fig.~\ref{fig:ellipLoopsPlot}.  Interstitial \aloops{} are less elliptical than those of vacancy type \cite{jostsons1977nature}.  Conversely, TEM images show \cploops{} to be circular \cite{harte2017effect}.  The elliptical shape of \aloops{} indicates that the dislocation line energy is anisotropic.  This could be due to the anisotropic elastic behaviour of $\alpha$-Zr (i.e. a difference in the long-range displacement field), an anisotropy in the dislocation core structure (i.e. a difference in the short-range atomic arrangement), or both.  The fact that \cploops{} are circular is consistent with the symmetry in the basal plane.

Along with the effects of elastic and crystal anisotropy, anisotropic diffusion of point defects may have an influence on loop ellipticity.  A computational study by Fuse found that for Zr the stable interstitial site is the basal octahedral and that this diffuses preferentially in the basal plane \cite{fuse1985evaluation}.  A review by Frank \cite{frank1988intrinsic} of point defect studies, building on a review of computational work on the subject by Bacon \cite{bacon1988review}, reported this to be the general case for HCP metals with $c/a < \sqrt{8/3}$.  Frank's review also stated that the diffusion of vacancies is only weakly anisotropic \cite{frank1988intrinsic}, citing self-diffusion studies by Seeger and G{\"o}sele \cite{seeger1975vacancies} and Peterson \cite{peterson1978self}.  Molecular dynamics simulations by Osetsky et al.~\cite{osetsky2002anisotropy} and by Kulikov and Hou \cite{kulikov2005vacancy} found that, at reactor temperatures, interstitial diffusion occurs overwhelmingly in the basal plane.

Woo's 1988 theory utilises the diffusional anisotropy difference (DAD) of point defects to explain the existence of irradiation-induced defects seen in Zr and the associated IIG \cite{woo1988theory}.  It postulates that dislocation lines parallel to $[0001]$ capture more interstitials than vacancies, which makes the minor axis shorter for vacancy loops and the minor axis longer for interstitial loops \cite{woo1988theory}.  If vacancy and interstitial loops begin as equally elliptical, then the influence of DAD will be to make vacancy loops more elliptical and interstitial loops less elliptical  \cite{woo1988theory}.  Woo used a TEM study by Brimhall \cite{brimhall1971microstructural} to support the notion that elliptical loops have equal ellipticity without DAD \cite{woo1988theory}.  However Brimhall's study only identified one interstitial loop and the study was on irradiated $\alpha$-Ti \cite{brimhall1971microstructural}.  As $\alpha$-Ti is non-cubic, with a $c/a$ ratio less than ideal, it too should be affected by DAD \cite{wood1962lattice}.  Therefore the question of whether \aloops{} in Zr would be equally elliptical without DAD is unanswered, but is something we can probe by simulating elliptical loops in a model without diffusion effects.  Contradicting DAD is a recent Monte Carlo modelling study by Christiaen et al.~that found IIG could be replicated with diffusion behaviour where vacancies diffuse preferentially in the basal plane \cite{christiaen2020influence}.  Additionally, Christiaen et al.~reference ab intio studies of Zr that suggest vacancies diffuse more rapidly in the basal place, although they acknowledge that alloying elements could alter this \cite{christiaen2020influence}.\\

\subsection{Stacking Faults.}

\noindent A faulted dislocation loop contains a stacking fault and this contributes to the formation energy of the loop.  Stacking faults important to our present work are those on the basal plane $\{0001\}$, the 1st prismatic plane $\{10\bar{1}0\}$ and the 2nd prismatic plane $\{2\bar{1}\bar{1}0\}$.  On the basal plane three types of stacking fault are pertinent: high energy (HE), first intrinsic (I1) and extrinsic (E).  One mechanism for a \cploop{} to form is for it to begin as a vacancy platelet on the basal plane, formed by agglomerated irradiation-induced vacancies, which collapses to form a HE stacking fault.  This gives the HE stacking fault a plane sequence of `ABABBAB' (where the HCP plane sequence is denoted `ABABAB').  The loop grows by absorbing additional vacancies to the point where the HE stacking fault energy becomes so great that the loop shears to produce a lower energy I1 stacking fault, in order to reduce the loop's energy.  This shearing is achieved by a partial dislocation with burgers vector $1/3\langle1 \bar{1} 0 0\rangle$ sweeping across the HE stacking fault \cite{hull2001introduction}.  The HE stacking fault is bounded by a dislocation line with burgers vector: $\bold{b}_{\ttm{HE}} = 1/2[0001]$ and we term this type of loop a \chloop{}.  The combination of the dislocation bounding the HE stacking fault with the partial dislocation results in a burgers vector of $1/6\langle2 \bar{2} 0 3\rangle$ via the reaction

\begin{equation}\label{eq:gradient}
\frac{1}{2}[0001] + \frac{1}{3}\langle1 \bar{1} 0 0\rangle =\frac{1}{6}\langle2 \bar{2} 0 3\rangle, 
\end{equation}
	
\noindent which is the burgers vector, $\bold{b}_{\ttm{I1}}$, of the dislocation line bounding the I1 stacking fault.  This transition changes a \chloop{} into a \cploop{}.  For Zr containing an I1 stacking fault, the plane sequence is `ABABCBC', meaning that it is a HCP structure containing a layer with face centred cubic (FCC) structure.  

A $c$-component loop with $\bold{b}_{\ttm{HE}} = 1/2[0001]$ but containing an extrinsic stacking fault is also of interest and we term this a $\langle c/2 \rangle_{\ttm{EXT}}$ loop.  As an extrinsic fault has lower energy than a HE stacking fault, transformation from a \chloop{} to a $\langle c/2 \rangle_{\ttm{EXT}}$ loop may occur followed by a transition to a \cploop{}, rather than a direct transition from a \chloop{} to a \cploop{} \cite{hull2001introduction}.\

A further type of $c$-component loop potentially exists, the \cloop{}, which is a perfect loop containing no stacking fault.  We expect that above some threshold radius, a \cloop{} will have lower energy than a \cploop{} containing the same number of vacancies, because although the burgers vector magnitude of the \cloop{} is greater, it is truly unfaulted, whereas the \cploop{} still contains an I1 fault.  Griffths and Gilbert observed \cloops{} with TEM, in irradiated and annealed Zircaloy-4 \cite{griffiths1987formation}.  Griffths and Gilbert stated that the formation of \cloops{} occurs via the reaction
\begin{equation}\label{eq:doubleClp}
\frac{1}{6}\langle2 0 \bar{2} \bar{3}\rangle + \frac{1}{6}\langle\bar{2} 0 2 \bar{3}\rangle = [0 0 0 \bar{1}],
\end{equation}
	
\noindent in which two \cploops{} coalesce to form a \cloop{} \cite{griffiths1987formation}.  \cloops{} have also been observed in irradiated Mg, which like Zr has a HCP crystal structure \cite{xu2017origin}.  Despite reports of their existence, \cloops{} have been little studied.\

An unfaulting process would occur for an \aloop{}, if it initially forms on a $\{1 \bar{1} 0 0\}$ plane as a faulted prismatic loop with burgers vector $\frac{1}{2}\langle0 1 \bar{1} 0\rangle$.  This faulted prismatic \aloop{} could unfault to form a sheared \aloop{} with the observed burgers vector of $\frac{1}{3} \langle1 1 \bar{2} 0\rangle$ \cite{varvenne2014vacancy}.  The determination of unfaulting radii for \aloops{} and \chloops{} is important for improved understanding of nascent loops, where we define a nascent loop as one with a diameter less than 10~nm.  Varvenne et al.~performed a computational study in which they determined the unfaulting radius for faulted vacancy \aloops{} to be 2.7\ nm and the unfaulting radius for \chloops{} to be 2.4\ nm \cite{varvenne2014vacancy}.\\

\subsection{Dislocation Loop Nucleation.}
\noindent Nucleation of dislocation loops is an area where there remains a great deal of uncertainty.  It is believed that loop nucleation begins with the agglomeration of irradiation-induced point defects into clusters \cite{averback1987energetic,woo1990diffusion}.  The clusters may then absorb more point defects to the point where they collapse to create a dislocation loop \cite{woo2009generation}.   Experimentally \aloops{} are observed early on in the irradiation process, whilst $c$-component loops do not appear until higher doses \cite{aHarteThesis}.

\begin{figure}[!ht]\begin{center}
{\includegraphics[width=\figwidth]{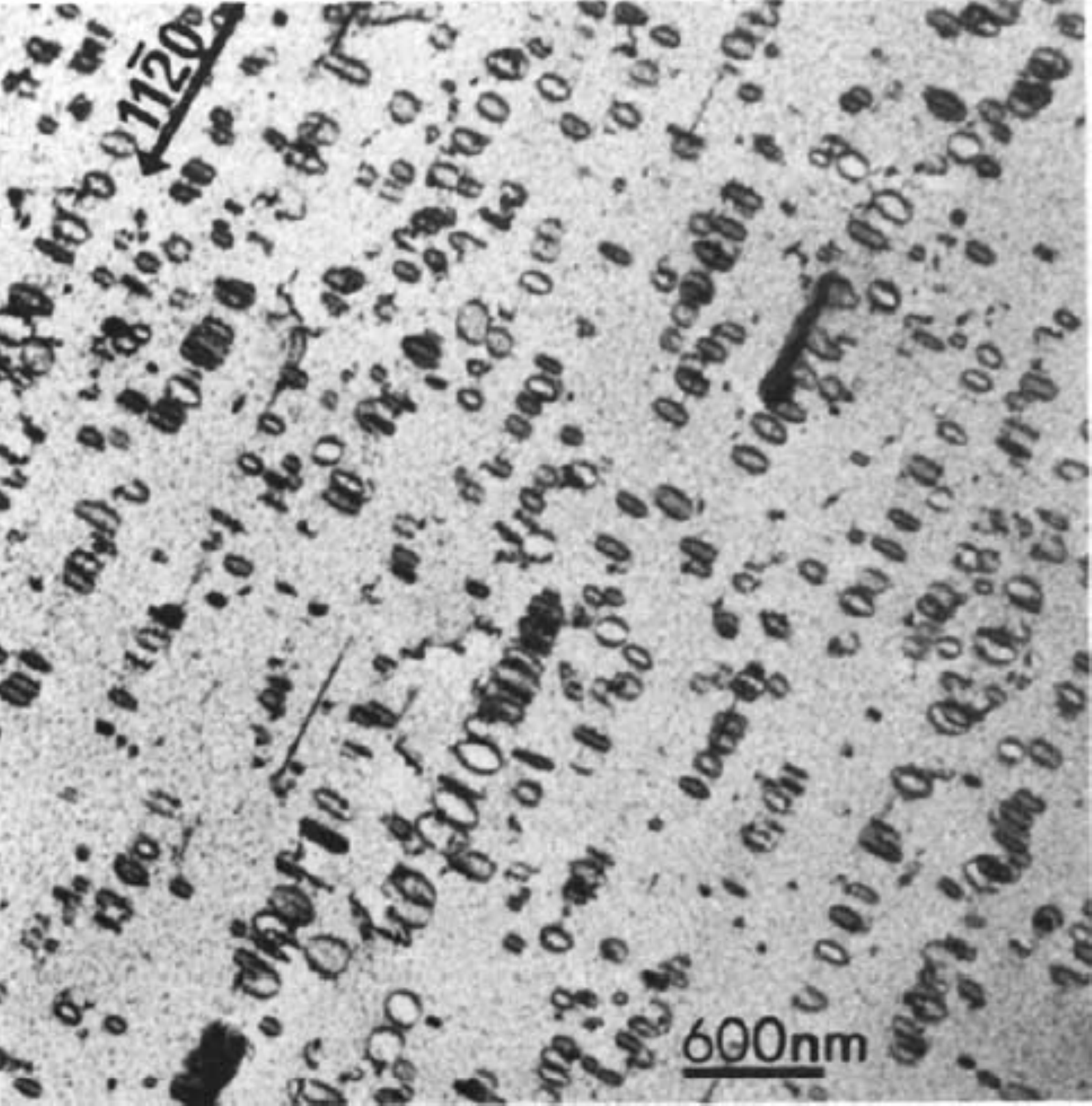}}	%
\caption{The elliptical shape of \aloops{} can be seen in this TEM image of irradiated zone refined Zr.  Ordering of the \aloops{} along the trace of the basal plane is also evident (Reproduced from Jostsons et al.~\cite{jostsons1977nature}, with permission from Elsevier.).}
\label{fig:aLoopsJostsons}
\end{center}
\end{figure}

Naturally, multiple loops will nucleate and orient themselves in a way that reduces their strain field interaction.  As can be seen in Fig.~\ref{fig:aLoopsJostsons}, \aloops{} arrange themselves along the trace of the basal plane.  The TEM image seen in Fig.~\ref{fig:aLoopsJostsons} is a 2D image of a Zr crystal oriented with the [$1 \bar{1} 0 0$] normal to the page.  Hence the image shows many ($1 \bar{1} 0 0$) plane loops throughout the depth of the TEM foil.  Therefore Fig.~\ref{fig:aLoopsJostsons} shows the \aloop{} positions in the image plane along [0001] and $[11\bar{2}0]$, but the positions along [$1\bar{1}00$] cannot be determined and so the degree of ordering along [$1 \bar{1} 0 0$] is unknown.  The ordering of loop types has not been entirely established experimentally.  Whilst it might be expected that ordered loops alternate between interstitial and vacancy character to minimise the interaction of their strain fields, to our knowledge this has not been confirmed.\  %

	The computational atomistic simulation study reported in this paper addresses the phenomena outlined above.  Whilst TEM studies have done much to reveal details about dislocation loop structure at the nanoscale, computer models allow inspection of individual atoms, with none of the obscuring effects of experimental sample artefacts or limitations to atomic resolution.  Experimentally, it is difficult to resolve defects smaller than $\sim$5~nm using TEM \cite{jenkins1999application} and so simulations provide an advantage in this regard.  Defect types, sizes and configurations can be controlled in the simulation and their energies calculated.  This makes study of defect energetics accessible in a way that would be difficult experimentally.  These results are considered within the limitations of the model: the applicability of the potential employed, the use of pure Zr and the limited size of the simulated volume.

\section{Method}		%

\noindent To study the phenomena introduced in Section~\ref{sec:intro}, we have made use of atomistic models of bulk $\alpha$-Zr, containing carefully controlled defect populations.  The defect types studied are tabulated in Table \ref{loopsTable}.\

The simulation supercells were relaxed, and in some cases annealed, with the LAMMPS molecular modelling software package (\url{http://lammps.sandia.gov}) \cite{plimpton1995fast} using Mendelev \& Ackland's \#3 (M\&A \#3) potential.  This was selected because it closely reproduces the stacking fault energies (SFE) of Zr, which was a deficiency of previous Zr potentials \cite{mendelev2007development}.  As this study is concerned with dislocation loops, which usually contain a faulted plane, the ability of a potential to capture SFE is crucial.  M\&A \#3 also replicates interstitial formation energies well, with the octahedral (O) site being the most stable, followed by the BO.  The formation energies were close, with the former being 2.88~eV and the latter 2.90~eV in good agreement with \textit{ab initio} results in the literature \cite{de2011structure}.  Some more recent \textit{ab initio} results by Peng et al.~suggest that BO is more stable than O, (formation energies 2.78~eV and 2.92~eV respectively) \cite{peng2012stability},  although the absolute formation energies are still close to those determined by Peng et al.\ and replicated by M\&A \#3. 

The OVITO visualisation software package \cite{stukowski2009visualization} was used to analyse the relaxed supercell configurations. \
\if
 To check the suitability of M\&A \#3 for modelling dislocation lines we carried out a number of density functional theory (DFT) benchmarking calculations.  DFT calculations were carried out on a series of infinite dislocation dipoles in the $[11\bar{2}0]$ direction and a series of infinite dislocation dipoles in the [0001] direction.  The DFT calculations were compared with the equivalent calculations performed with LAMMPS using  M\&A \#3.  For the $[11\bar{2}0]$ direction dislocations the line energy agreement is unexpectedly good: within a few \% of the DFT results.  For the [0001] direction dislocations the agreement is not as good and the difference between the LAMMPS and DFT calculated line energies is $\sim 45\%$.\\
\fi
\vspace{.5cm}
\begin{table*}[t!]
\begin{center}
\begin{tabular}{|c|c|c|c|c|}
\hline
          Loop type  &   Loop Character & Habit Plane.  &  Burgers Vector     &  Burgers Vector             \\
                      &    Studied &   &       &    Magnitude (Ang)            \\[1.ex]
    \hline
&     &   &       &          \\[-1.ex]
${\langle c/2+p \rangle}$              & Vacancy and SIA  & $ (0001) $ &  $\frac{1}{6} \langle 2 \bar{2} 0 3 \rangle$   &		3.18 	  	               \\ [1.ex]
${\langle c/2 \rangle}$              & Vacancy  & $(0001) $ &  $\langle 0 0 0 2 \rangle$   &		2.57 	                \\[1.ex]
$\langle c/2 \rangle_{\ttm{EXT}}$              & Vacancy  & $(0001) $ &  $\langle 0 0 0 2 \rangle$   &		2.57 	                \\[1.ex]
${\langle c \rangle}$              & Vacancy  & $(0001)$ &  $\langle 0 0 0 1 \rangle$   &		5.15 	               \\ [1.ex]
$\langle a \rangle^{\text{1ord}}$              & Vacancy and SIA & $(0 1 \bar{1} 0)$ &  $\frac{1}{3} \langle 2 \bar{1} \bar{1} 0 \rangle$   &		3.23 	               \\[1.ex]
Edge $\langle a \rangle^{\text{1ord}}$              & Vacancy & $(0 1 \bar{1} 0)$ &  $\frac{1}{2} \langle 0 {1} \bar{1} 0 \rangle$   &		2.80 	               \\[1.ex]
$\langle a \rangle^{\text{2ord}}$              & Vacancy and SIA & $(1 \bar{2} 1 0)$ &  $\frac{1}{3} \langle 2 \bar{1} \bar{1} 0 \rangle$   &		3.23 	               \\ [1ex]
    \hline 
\end{tabular}
\captionsetup{width=.99\textwidth} %
\caption{The various dislocation loops included in this study. A $\langle c/2 \rangle$ loop contains a HE stacking fault and a $\langle c/2 \rangle_{\ttm{EXT}}$ loop contains an extrinsic stacking fault.}%
\label{loopsTable}
\end{center}
\end{table*}

\subsection{Dislocation Loop Construction.}
\noindent Dislocation loops were created by first removing or adding a platelet of atoms, depending on the character of the loop.  The surrounding atoms were then displaced according to a model displacement field $\textbf{u}_0(\textbf{r})$.

Lazar and Kitchener \cite{lazar2013dislocation} provide a method to calculate the displacement field for a dislocation loop in an anisotropic medium based on continuum elasticity theory:
 \begin{equation}\label{eq:anisLoopDisp}
\bold{u}(\bold{r}) = - \frac{\bold{b} \Omega(\bold{r})}{4 \pi} + \bold{u}(\bold{r})_{\textrm{elastic}}.
\end{equation}
\noindent Here the first term is the plastic displacement and the second term is the elastic displacement \cite{lazar2013dislocation}.  $\Omega(\bold{r})$ is the solid angle subtended by the loop at point $\bold{r}$ and the loop's burgers vector is $\bold{b}$.  In our study we are considering elliptical dislocation loops and since the calculation of the solid angle subtended by an ellipse at a general point is non-trivial we have adopted an alternative, pragmatic approach.
We use a model displacement field $\textbf{u}_0(\textbf{r})$, and then rely on relaxation under the model interatomic forces to displace the atoms into the final relaxed form, $\textbf{u}(\textbf{r})$. We choose the form
\begin{equation}
\textbf{u}_0(\textbf{r}) = \bold{b} ~\alpha(\mu) \beta({d}),
\end{equation}
where \textbf{r} is the initial position of an atom, \bold{b} is the burgers vector of the loop and $\alpha(\mu)$ and $\beta(d)$ are two functions that determine the rate of decay of the displacement field in directions normal to and parallel to the habit plane of the loop. These functions are defined in detail in Appendix~\ref{app:dislocationconstruction}.

\subsection{Energy minimisation and selecting dislocation loop densities.}\label{SelectDislocDens_pap1}
\noindent To arrive at the stable structure for our dislocations, we relax the atomic coordinates using the Polak-Ribiere version of the conjugate gradient (CG) minimisation algorithm with an energy tolerance of $10^{-10}$ (unitless) and a force tolerance of $10^{-6}$~eV/{\AA}. To ensure that we reach a valid final state, we run a series of five minimisations alternating between relaxing only the atomic positions within a fixed supercell and also relaxing the supercell under conditions of the CG minimisation algorithm using the same tolerances as previously.

We apply full periodic boundary conditions to simulate infinite bulk material. We also make use of skewed supercells to avoid the case where the periodic copies of the dislocation loops are stacked directly on top of one another, which would give rise to particularly unrealistic strain interactions. Ascertaining whether loops in reality are offset in this way is difficult.  TEM micrographs, such as Fig.~\ref{fig:aLoopsJostsons}, reveal that \aloops{} are ordered in rafts.  However, as micrographs are 2-D projections it is not possible to see whether the loops are configured in an ordered manner along $\langle1 \bar{1} 0 0\rangle$.  Nonetheless, for loops of like character it is reasonable to assume that loops nucleate in a staggered manner, to minimise their strain interaction, and this arrangement is replicated by the skewing of the supercell.\

\begin{figure}[htbp!]\begin{center}
{\includegraphics[width=\figwidth]{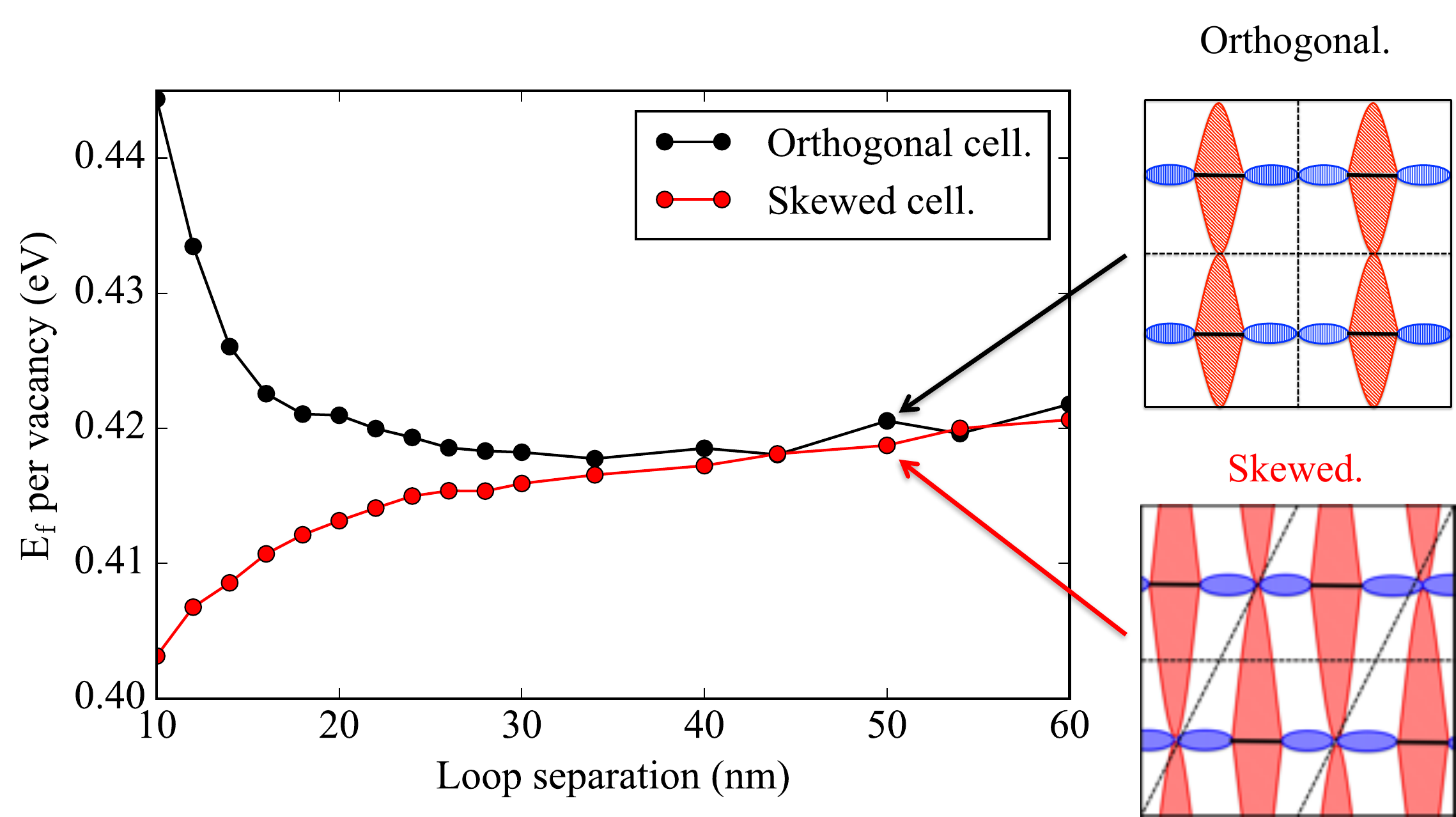}}. %
\caption{Formation energy per vacancy of a 10~nm \aloop{} as a function of separation from its periodic image.  The schematic illustrations show the strain field interactions, with the red shaded area representing tensile strain and the blue compressive strain.}
\label{fig:aLpSeparation}
\end{center}
\end{figure}

Excessive elastic interaction between a loop and its periodic images will lead to erroneously high strain interactions and could give rise to unrealistic structures. Thus we relaxed a test series of loops, surrounded by a varying thickness of bulk Zr padding, producing different effective separation distances, the results of which are shown in Fig.~\ref{fig:aLpSeparation}.  Kulikov and Hou \cite{kulikov2005vacancy} took such defect interaction into account in a 2005 study on loop energies in Zr, maximally separating defects across periodic boundaries.\ %
For skewed simulation boxes the formation energy, E$_\text{f}$, per vacancy increases with loop separation because the overlapping strain fields are of dissimilar character.  For this reason, whereas the orthogonal box gives a repulsive interaction, the skewed box gives an attractive one.  This is illustrated schematically in Fig.~\ref{fig:aLpSeparation} which shows how like strain fields overlap at relatively short range in the orthogonal box. In contrast, in the skewed cell the overlapping of the strain fields is only at longer range.  As loop separation increases, in larger simulation supercells, skewing has a reduced effect.

Based on the results shown in Fig.~\ref{fig:aLpSeparation} we chose a loop separation of 40~nm as sufficient to avoid undue strain interactions, without making the cells overly large.  We note that this separation creates a loop number density of around $10^{21}~$m$^{-3}$, comparable with that observed experimentally and that a separation of $\sim$50~nm was seen in TEM for c-component loops \cite{harte2017effect}.
\section{Results}\label{sec:results}	%
\subsection{Nascent Loop Transformations.}\label{sec:nascLoops}	

\subsubsection{Transformations of c-component loops.}\label{cLpTrans}
\noindent 
We begin by examining small, nascent loops, below the size easily resolved experimentally. These are of interest because they could reveal details of the nucleation process and the transition from one growth stage to another.  
In an atomistic modelling study de Diego et al.~\cite{de2008structure} simulated vacancy clusters on the basal plane that transformed into \chloops{}.  This supports the hypothesis that c-component loops begin as a cluster of vacancies, which then collapses to form a platelet.  As a \chloop{} absorbs more vacancies there is an increased impetus for it to shear to a lower energy configuration: a \cploop{}.  

The radius at which this happens, $r^*_{\text{HE}\rightarrow \text{I1}}$, was recently determined by Varvenne et al.~to be 2.4~nm. Varvenne et al.~used the M\&A \#2 potential, which better describes vacancy cluster energies \cite{mendelev2007development,varvenne2014vacancy}, whereas we have adopted M\&A \#3 potential for an improved representation stacking faults. We have therefore repeated the calculation of $r^*_{\text{HE}\rightarrow \text{I1}}$ by creating a series of \chloops{} and a series of \cploops{}, of varying diameter, contained in 140~nm x 140~nm x 40~nm supercells that house around 33 million atoms.  We relaxed these, analysed the energies and modelled the loop energy, $E(r)$, as the sum of energy contributions from the stacking fault and the bounding dislocation line:

\begin{equation}\label{eq:loopE}
E(r) = \gamma ~\pi (r-\Delta r)^2 + \lambda ~2 \pi (r-\Delta r). 
\end{equation}

Here $\gamma$ is the loop stacking fault energy per unit area and $\lambda$ is the energy per unit line length of the bounding dislocation. We treated $\gamma$ and $\lambda$ as fitting constants for the series of simulation results for $E(r)$.  $\Delta r$ is introduced to account for a difference between the nominal loop radius, $r$, used in constructing the loops and an effective radius $(r-\Delta r)$, post relaxation, that emerges from the simulation results.

\begin{figure}[htbp!]\begin{center}
\begin{subfigure}[b]{\figwidth}
\includegraphics[width=\figwidth]{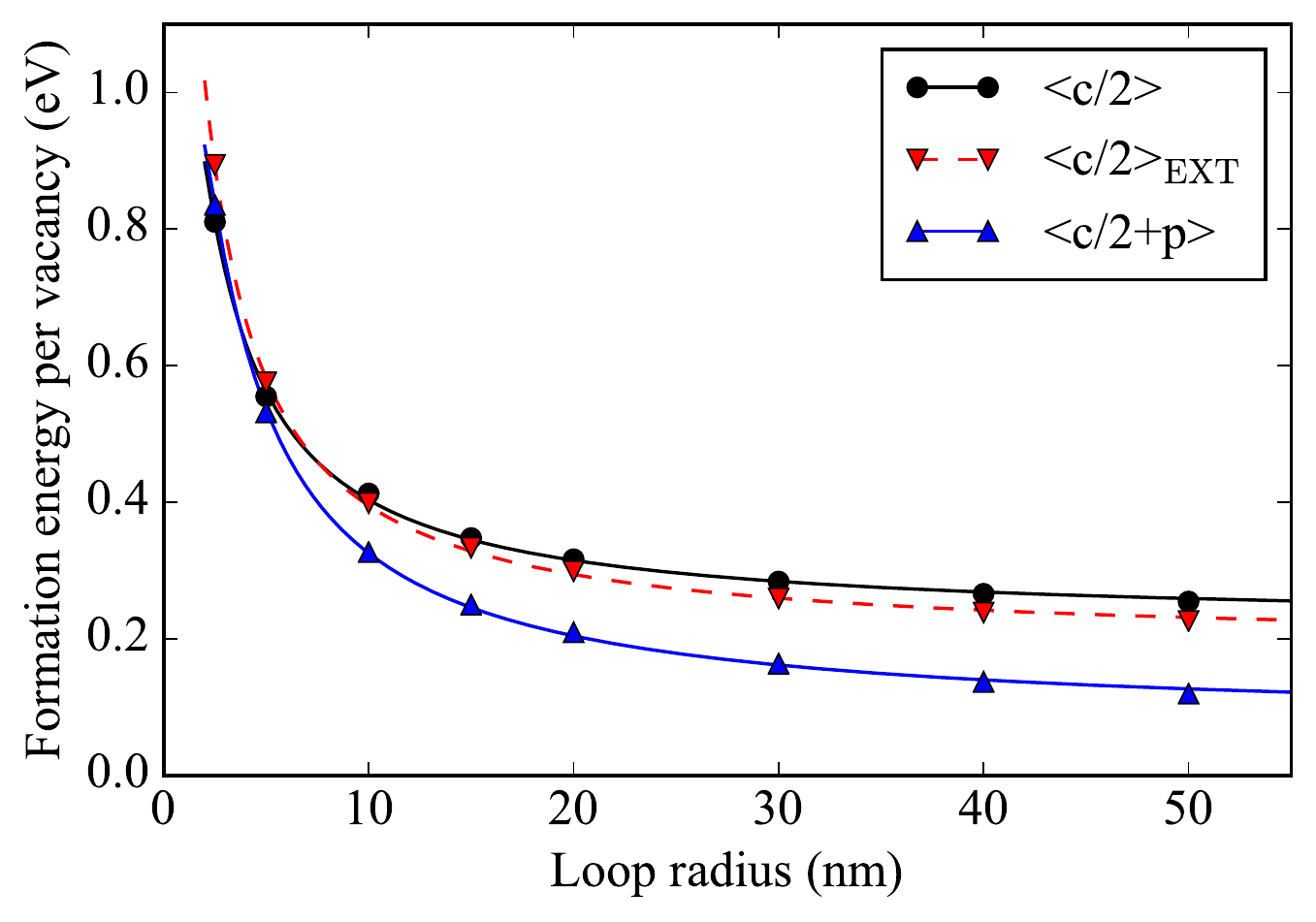} %
\captionsetup{justification=centering,singlelinecheck=false}  %
\caption{}
\end{subfigure}
\begin{subfigure}[b]{\figwidth}
\includegraphics[width=\figwidth]{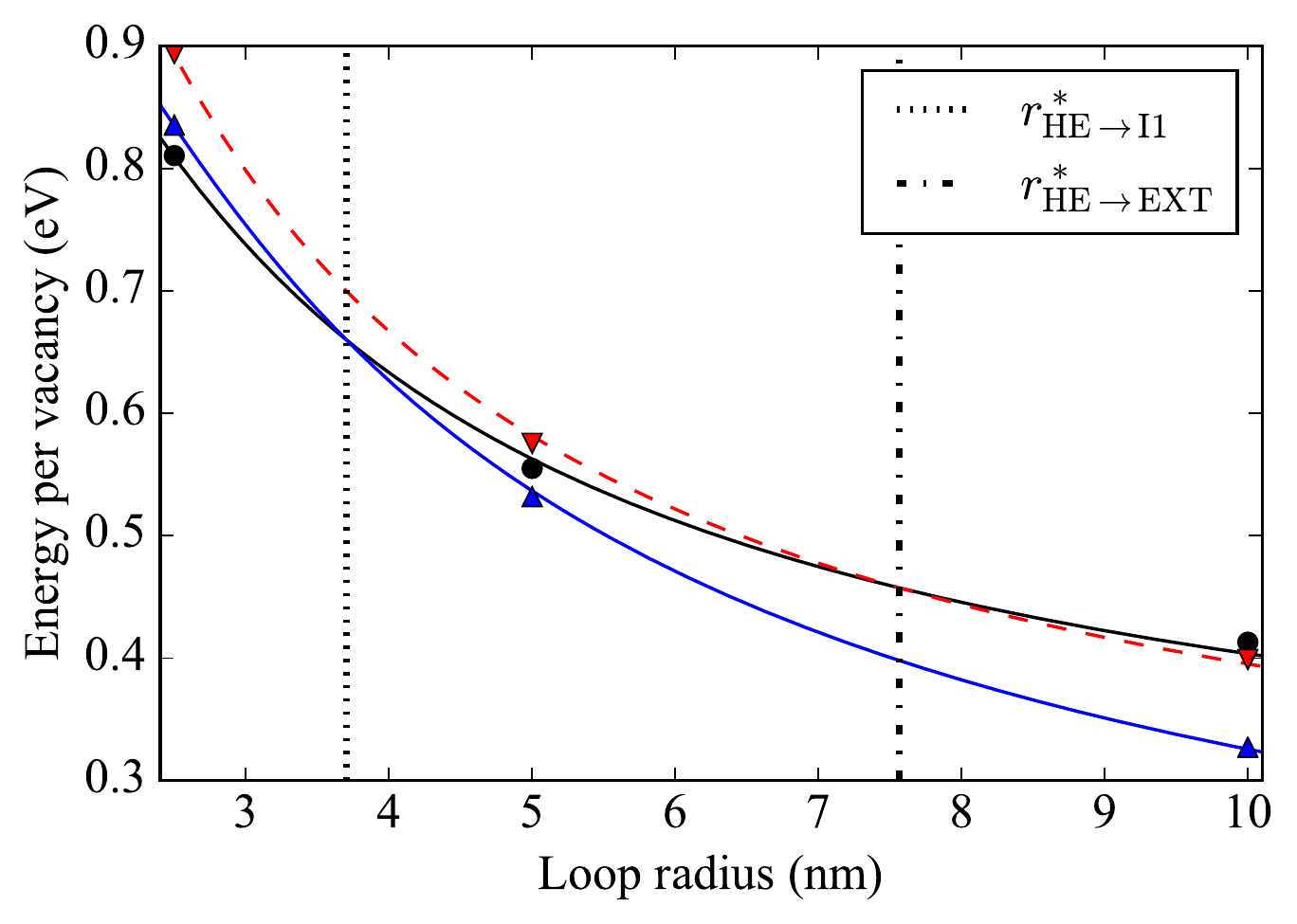} %
\captionsetup{justification=centering,singlelinecheck=false}  %
\caption{}
\end{subfigure}
\caption{The formation energy per vacancy as a function of radius for c-loops.  $\langle c/2+p \rangle$, $\langle c/2 \rangle_{\text{EXT}}$ and $\langle c/2 \rangle$ loop data are displayed in image (a), along with their fitted functions.  Markers are the simulation data and lines the fitted functions.  Image (b) shows the crossover radii, $r^*_{\mathrm{HE}\rightarrow \text{I1}}$ and $r^*_{\text{HE}\rightarrow \text{EXT}}$.}
\label{fig:cLpUnfault}
\end{center}
\end{figure}

The fitted functions for the $\langle c/2 \rangle$  series and the $\langle c/2+p \rangle$  series are plotted in Fig.~\ref{fig:cLpUnfault}.  The intersection of the functions gives $r^*_{\text{HE}\rightarrow \text{I1}} = 3.21$~nm, marked by the dashed line in Fig.~\ref{fig:cLpUnfault}.  Table \ref{sect1.1table} contains $r^*_{\text{HE}\rightarrow \text{I1}}$ along with $\gamma$ and $\lambda$ for \chloops{} and \cploops{}, along with some of these values from other sources.

Our fitted functions for $\langle c/2 \rangle_{\ttm{EXT}}$ and $\langle c/2 + p \rangle$ loops predict that the energy of the former will always be higher than that of the latter.  However, Varvenne et al.'s \cite{varvenne2014vacancy} study \emph{does} predict a transition from a $\langle c/2 \rangle_{\ttm{EXT}}$ loop to a \cploop{} at a radius of $r^*_{\text{E}\rightarrow \text{I1}}=$ 2.4~nm.  M\&A~\#3 is superior for SFEs, as shown in Table \ref{sect1.1table}, and for the transformation to an I1 stacking fault, the SFE is more important than the vacancy binding energies, as the number of vacancies is conserved during this transformation.  Additionally, the vacancies are obliterated when the platelet collapses and so can no longer be considered a cluster.  This was discussed by Varvenne at al., as they found that with M\&A~\#2 cavities were more stable than vacancy dislocation loops, but this is evidently not the case in reality as cavities are rarely seen in irradiated Zr \cite{varvenne2014vacancy,griffiths1988review}.

\begin{table*}[t!]
\begin{center}
\begin{tabular}{|c|c|c|c|c|c|}
\hline
       &   This study  &\multicolumn{2}{c}{} Varvenne \cite{varvenne2014vacancy} &&    Domain \cite{domain2004atomic}  \\ [1ex] \cline{3-5}%
	&   EAM & DFT & EAM & EAM  & \textit{Ab Initio}  \\ [1ex]
	&   M\&A \#3  &  & M\&A \#2 & M\&A \#3   &  \\ [1ex]
    \hline
$r^*_{\text{HE}\rightarrow \text{I1}}$~(nm)          & 3.70     &  &   &  & \\ [1ex]
$r^*_{\text{HE}\rightarrow \text{E}}$~(nm)          & 7.56     &  &  & &    \\ [1ex]
$r^*_{\text{E}\rightarrow \text{I1}}$~(nm)          & n/a     & 1.4 & 2.4  &&   \\ [1ex]
$r^*_{\text{I1}\rightarrow \text{Unfaulted}}$~(nm)          & 56.1 &  &   &&   \\ [1ex]
$\gamma_{\text{I1}}$~(mJ/m$^2$) &129 &  	147 	 & 	55        & 99 & 124       \\[1ex] %
$\gamma_{\text{EXT}}$~(mJ/m$^2$) & 333 &  	274 	 & 	164      & 297 & 249       \\[1ex] %
$\gamma_{\text{HE}}$~(mJ/m$^2$) &390 &&&&\\[1ex]%
$\gamma_{\text{SIA}}$~(mJ/m$^2$) &140 &&&&\\[1ex]%
$\lambda_{\text{c/2+p}}$~(eV/nm)               & 15.68 &   &         &  &                \\ [1ex]
$\lambda_{\text{c/2}}$~(eV/nm)		& 12.15 	&&&&	\\ [1ex]
$\lambda_{\text{SIA}}$~(eV/nm)		& 14.84  &&&&	\\ [1ex]
    \hline 
\end{tabular}
\caption{Results from this study are tabulated here, along with results from other computational studies.  EAM refers to the embedded atom method potential used in the atomistic modelling.  The value $r^*_{\text{HE}\rightarrow \text{I1}}$ is the radius at which a $\langle c/2 \rangle$ loop's energy becomes greater than that of an equivalent \cploop{}.  Similarly, $r^*_{\text{HE}\rightarrow \text{E}}$ is the radius for transormation from a \chloop{} to a $\langle c/2 \rangle_\text{EXT}$, $r^*_{\text{E}\rightarrow \text{I1}}$ is that from a $\langle c/2 \rangle_{\text{EXT}}$ loop to a $\langle c/2 + p \rangle$ and $r^*_{\text{I1}\rightarrow \text{Unfaulted}}$  that from a \cploop{} to a \cloop{}.  Our results show that at no radii do $\langle c/2 \rangle_{\text{EXT}}$ loop have lower energy than \cploops{} and n/a is used to denote this.}
\label{sect1.1table}
\end{center}
\end{table*}

\indent Our analysis implies a transition directly from the HE stacking fault to the I1, but Varvenne et al.~\cite{varvenne2014vacancy} also determined the radius at which the transition from an E stacking fault to an I1 occurs, implying a stacking fault transformation sequence HE $\rightarrow$ E $\rightarrow$ I1.  To explore this issue further we conducted a series of 800~K annealing simulations on \chloops{}, with radii from 0.5~nm to 10~nm, containing a HE stacking fault.  {Each simulation consisted of a 10~ps heating phase to 800~K, then a 20~ps anneal and a 15~ps cooling phase.  These were carried out in a simulation box at constant volume with a Nose-Hoover thermostat.  We found that for loops of 2.5~nm radius and below, the post annealing simulations contained defects resembling stacking fault tetrahedra, where no dislocation line could be discerned.  At radii of 3~nm and above the annealed cells contained discernible dislocation loops.  Analysis of these revealed that their burgers vector was $1/6 \langle20\bar{2}3\rangle$ and common neighbour analysis revealed the loop plane to be a single FCC layer.  This indicates that the loops with radii of 3~nm and above are \cploops{}.  The 3~nm radius \cploop{} was tilted towards the pyramidal plane and this concurs with observations of small, tilted c-component loops in TEM \cite{topping2018effect,griffiths1987neutron,griffiths1993hvem}.  The degree of tilting gradually decreased with increasing radius and almost no tilt was present at 10~nm.  These annealed loops suggest that the HE stacking fault configuration initially collapses to form a stacking fault tetrahedron, which then becomes a \cploop{}, following the absorption of more vacancies.  This suggests \chloops{}, containing either a HE or an extrinsic stacking fault, are not stable.  However, Griffiths observed loops with $\textbf{b} = 1/2[0001]$ on the basal plane, but these occurred after electron irradiation, which produces Frenkel pairs \cite{griffiths1984anisotropic}.  Neutron and proton irradiation produce collision cascades, which create large clusters for neutron irradiation and sequences of smaller clusters for proton irradiation \cite{was2017fundamentals}.  Therefore, the defects that electron irradiation produces may not be comparable to defects produced by these heavier particles.  Nonetheless, the observation of basal loops with $\textbf{b} = 1/2[0001]$ indicates that \chloops{} do exist.\

Whilst small discrepancies exist in the determination of the radius at which \cploops{} become energetically optimal, all the above results indicate that basal plane loops with radii above $\sim4$~nm are \cploops{}.  However, as impurities change SFEs, transformation radii may be altered in real alloys \cite{de2008structure}.  This highlights the necessity for future development of Zr alloy empirical potentials.\

%
%
\iffalse
\subsubsection{Discussion of Nascent Loop Transformations.}
%
%
%

\begin{figure}[htbp!]\begin{center}
\resizebox{2.5in}{!}{
%
\includegraphics[width=0.5\figwidth]{../images/5nmLoop/5nmAnneal.png}}
%
\\[1ex]
\resizebox{\figwidth}{!}{
\includegraphics[]{../images/6nmLoop/6nmLoopAnneal.png}}
\\[1ex]
\resizebox{2.5in}{!}{
\includegraphics[]{../images/10nmLoop/10nmAnnealFront.png}}
\\[1ex]
\resizebox{2.5in}{!}{
\includegraphics[]{../images/20nmLoop/anneal20nm15nmPad.png}}

\caption{Annealed cells that initially contained a $\langle c/2 \rangle$ with a HE stacking fault, are shown here.  From top to bottom the initial loop radii are: 2.5~nm, 3~nm, 5~nm and 10~nm.  Post-anneal the 2.5~nm loop has transformed to what appears to be a stacking fault tetrahedron.   The red lines present in the 3~nm and 10~nm loop images are the bounding dislocation line, which is absent in the 2.5 nm image as no discernible dislocation line was present.  The dislocation loop is highly tilted for the 3~nm loop and only marginally tilted for the 10~nm loop.  Although the dislocation loop was tilted the faulted loop plane remained in the basal plane.\dc{This image will be refined later}}
%
\end{center}
\end{figure}
\dc{Small (3nm) loops looks hexagonal and big (>10nm) circular. In Varvenne 2014 they built hex c-loops.  Perhaps we should include some words about c-loop shape.}
%
\fi
%
\subsubsection{Habit Plane of $\langle a \rangle$ loops.}\label{aLoopPlane}
\noindent Studies by TEM of irradiated zirconium have observed \aloops{} with a variety of loop normals from $[10\bar{10}]$ to $[11\bar{2}0]$, corresponding to loops on the 1st prismatic ${\{10\bar{1}0\}}$ and 2nd prismatic ${\{11\bar{2}0\}}$ planes respectively \cite{kelly1973characterization,jostsons1977nature}.  Although experimental evidence \cite{jostsons1977nature,northwood1979characterization,kelly1973characterization} (and see Fig.~\ref{fig:aLoopPlanes}) has shown the majority of the \aloop{} habit planes clustered around 1st prismatic planes, it is unclear why this preference for \aloops{} to inhabit ${\{10\bar{1}0\}}$ exists.  To address this we calculated the energies of \aloops{} with various diameters, for a series on $(0 1 \bar{1} 0)$ and for a series on $(1\bar{2}10)$.  The loops energies were fitted to a function as in Section \ref{cLpTrans}.  Loop series were created for both vacancy and interstitial \aloops{}, and these results are displayed in Fig.~\ref{fig:aLp1prisV2pris} and tabulated in Table~\ref{aLpEtable}.

\begin{table}[h!]
\begin{center}
\begin{tabular}{|c|c|c|}
\hline
Defect  & $\lambda$~(eV/nm) & $\gamma$~(mJ/m$^2$)       \\ [1ex] %
    \hline
   &      &            \\     [-2.ex]
$\langle a \rangle^{\text{1ord}}_{\text{vac}}$   & 14.9      & 32           \\ 
Edge $\langle a \rangle^{\text{1ord}}_{\text{vac}}$   & 13.1      & 176           \\ 
$\langle a \rangle^{\text{2ord}}_{\text{vac}}$   & 15.6      & 39             \\ 
$\langle a \rangle^{\text{1ord}}_{\text{sia}}$	 &	13.7 & 49 		\\ 
$\langle a \rangle^{\text{2ord}}_{\text{sia}}$	 &	14.9 & 44 		\\ [1ex]
    \hline 
\end{tabular}
\caption{The $\langle a \rangle$ loop energetics results are tabulated here.}
\label{aLpEtable}
\end{center}
\end{table}

The interplay between the dislocation line energy density $\lambda$ and the stacking fault energy density $\gamma$ in determining which prismatic plane results in the lowest energy for an \aloop{} is complex.  This is because the effective area densities of point defects in \aloops{} on the 1st prismatic and 2nd prismatic planes are different. Analysis of \aloops{} on different prismatic planes must thus be in terms of defects accommodated within each loop, rather than loop radii.
The area density of the 1st prismatic plane is less than that of the 2nd prismatic plane, so for \aloops{} containing an equivalent number of defects, the \aloop{} on the 1st prismatic plane will have a larger radius than that on the 2nd, meaning it has greater bounding dislocation line length.  However, $\lambda$ also varies with habit plane and the overall dislocation line energy will depend on both $\lambda$ and the circumference. The line energy is anisotropic, but in this case we model only circular loops and treat $\lambda$ as a mean value.

\begin{figure}[]\begin{center}
\begin{subfigure}[b]{\figwidth}
\includegraphics[width=\figwidth]{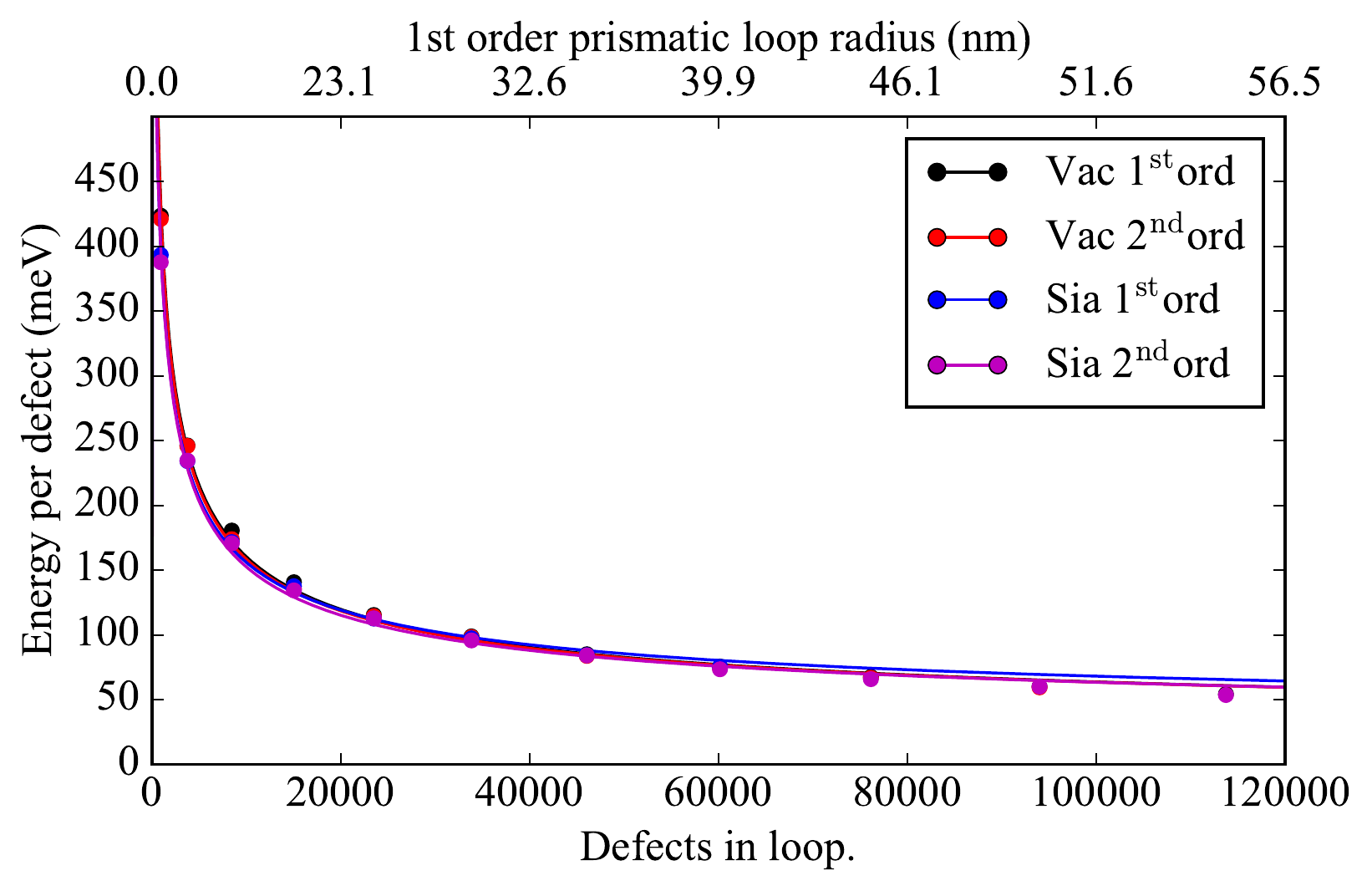}%
\captionsetup{justification=centering,singlelinecheck=false}  %
\caption{}
\end{subfigure}
\begin{subfigure}[b]{0.75\figwidth}
\includegraphics[width=0.75\figwidth]{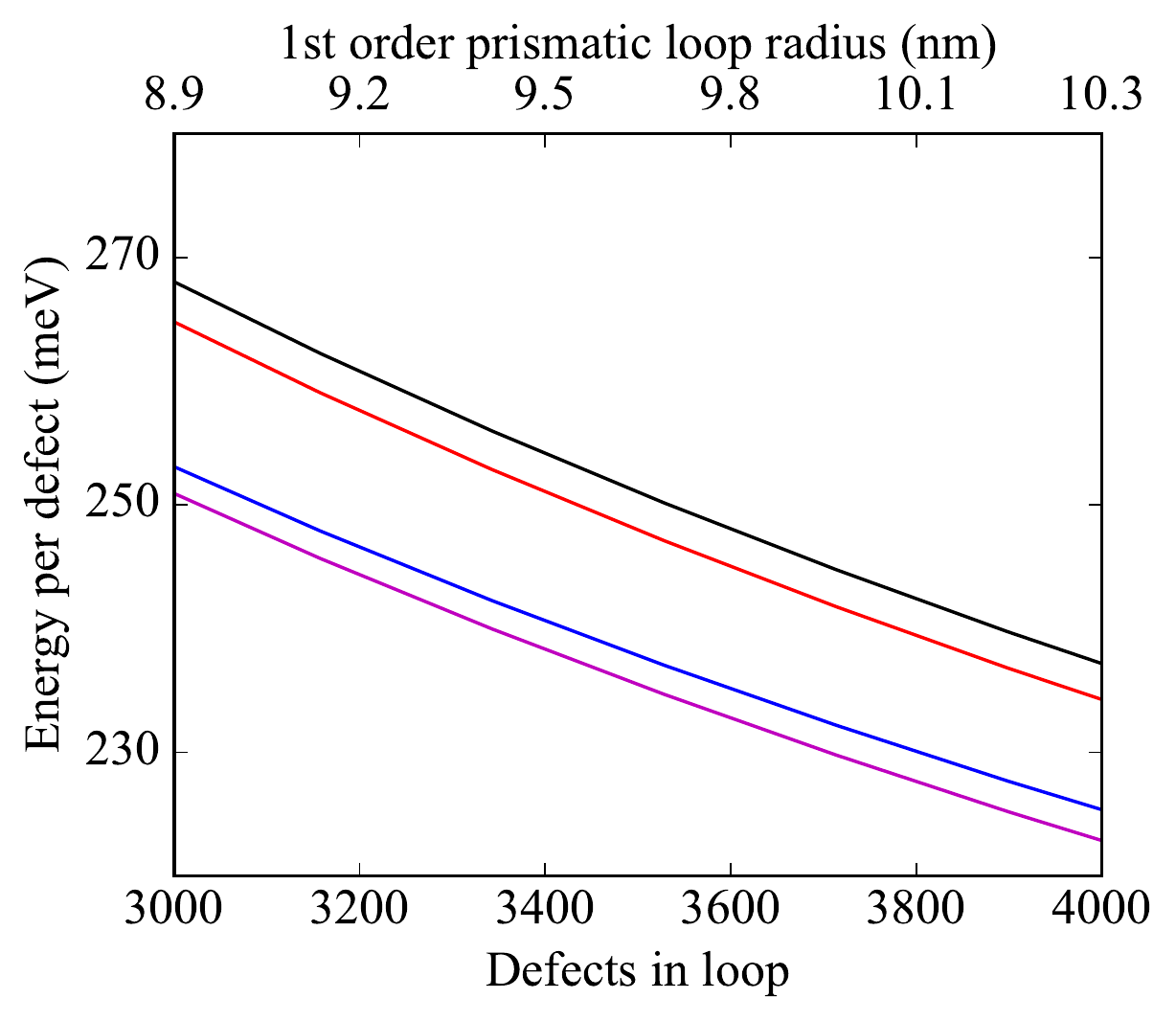}%
\captionsetup{justification=centering,singlelinecheck=false}  %
\caption{}
\end{subfigure}
\begin{subfigure}[b]{0.75\figwidth}
\includegraphics[width=0.75\figwidth]{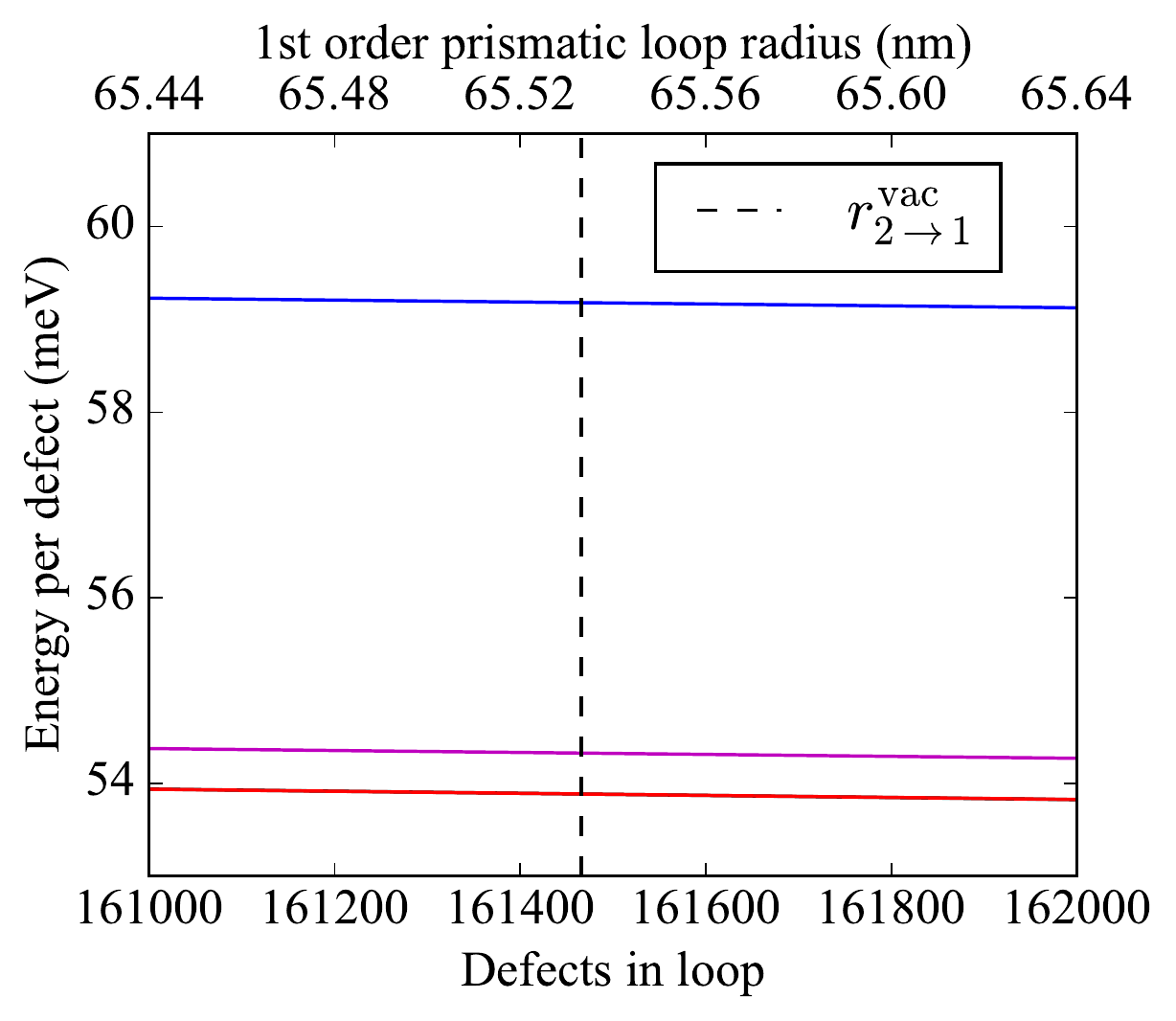}%
\captionsetup{justification=centering,singlelinecheck=false}  %
\caption{}
\end{subfigure}
\caption{Formation energy per defect as a function of radius, for SIA and vacancy $\langle a \rangle$ loops inhabiting the 1st and 2nd prismatic planes.  Circles represent the simulation results and lines are the model fit.  Image (a) shows the series across the full range of radii studied.  Only the data points for one set are visible in the image (a) because the points for the other 3 sets effectively lie in the same position, for the scale used.  Note that in image (c) the `Vac 1$^{\text{st}}$ ord' line is concealed by the overlying `Vac 2$^{\text{nd}}$ ord' line.}
\label{fig:aLp1prisV2pris}
\end{center}
\end{figure}

As sheared $\langle a \rangle$ loops are unfaulted, we would expect the fitted $\gamma$ values to be zero.  However, in all cases $\gamma$ has a small but finite value (around 25\% of $\gamma_\text{I1}$) that we believe is due to minor dis-registries of atoms due to the strain field propagating out from the dislocation core.  As $\gamma$ is non-zero, the surface defect energy will dominate the line energy for large loops, but as the $\gamma : \lambda$ ratio is small in comparison to c-component loops, this happens at much larger sizes than was the case for c-component loops.  This is demonstrated in Fig.~\ref{fig:aLp1prisV2pris}(b) and (c) which show small and large loop energies respectively.  Small $\langle a \rangle^{\text{1ord}}_{\text{vac}}$ loops have higher energy than $\langle a \rangle^{\text{2ord}}_{\text{vac}}$ loops, but at larger radius this difference diminishes and at radii above $\sim 65$~nm the order reverses.  At all radii $\langle a \rangle^{\text{1ord}}_{\text{sia}}$ loops have higher energy than $\langle a \rangle^{\text{2ord}}_{\text{sia}}$ loops and this difference increases with radius.\

In all cases, the differences in energy per defect of the various \aloops{} are small: around several meV per defect.  Hence the thermodynamic driving force available for a loop transformation to occur is small, in comparison to that for c-component loops.  This may be the reason for the distribution of loop habit planes seen in Jostsons's work \cite{jostsons1977nature}.  For the size range of the loops that Jostsons was examining, the 2nd order prismatic habit plane results in the lowest energies, for both interstitial and vacancy \aloops{}.  If loops nucleate on the 1st prismatic, there may be a reduction in energy from rotating onto the 2nd prismatic, but the thermodynamic driving force may be insufficient for all loops to do so, as obstacles or large transformation energy barriers may impede this rotation.  For the loop radii observed in Jostsons et al.'s study of \aloop{} habit planes \cite{jostsons1977nature}, the energy reduction between 1st prismatic and 2nd prismatic habit planes is greater for interstitial loops.  This concurs with Jostsons et al.'s data, replicated in Fig.~\ref{fig:aLoopPlanes} in this paper, as more of the interstitial loops are distributed towards the 2nd prismatic plane than for the vacancy loops.
\subsubsection{Unfaulting radius of \aloops{}.}  %

\noindent Jostsons et al.~observed \aloops{} with TEM and, whether on the 1st or 2nd prismatic planes, saw them to have a burgers vector of $1/3\langle11\bar{2}0\rangle$ \cite{jostsons1977nature}.
Griffiths assumed that after nucleation, $\langle a \rangle$ loops have $\bold{b}=1/2\langle10\bar{1}0\rangle$ and then unfault via the reaction \noindent \cite{griffiths1991microstructure}

\begin{equation}\label{eq:aLpUfault}
\frac{1}{2}\langle10\bar{1}0\rangle + \frac{1}{6}\langle\bar{1}2\bar{1}0\rangle  \rightarrow \frac{1}{3}\langle11\bar{2}0\rangle.
\end{equation}

de Diego et al.~gave credence to this process, when they conducted an atomistic modelling study \cite{de2008structure} and observed the unfaulting of a vacancy dislocation loop, inhabiting a 1st prismatic plane, after a 150 ps anneal.  A defect platelet on the 1st prismatic plane that collapsed into a pure edge \aloop{} would have $\bold{b}=1/2\langle10\bar{1}0\rangle$ and a defect platelet on the 2nd prismatic plane that collapsed into a pure edge \aloop{} would have $\bold{b}=1/3\langle11\bar{2}0\rangle$.  As the magnitude of $\bold{b}=1/2\langle10\bar{1}0\rangle$ is lower than that of $\bold{b}=1/3\langle11\bar{2}0\rangle$ we can expect that $\lambda$ resulting from the former will be lower than that of the latter, in accordance with Frank's rule.  Line energy dominates the total energy at small radii, hence we presume that \aloops{} nucleate on the 1st prismatic plane.  

\begin{figure}[htbp!]\begin{center}
\begin{subfigure}[b]{\figwidth}
\centering
\includegraphics[width=\figwidth]{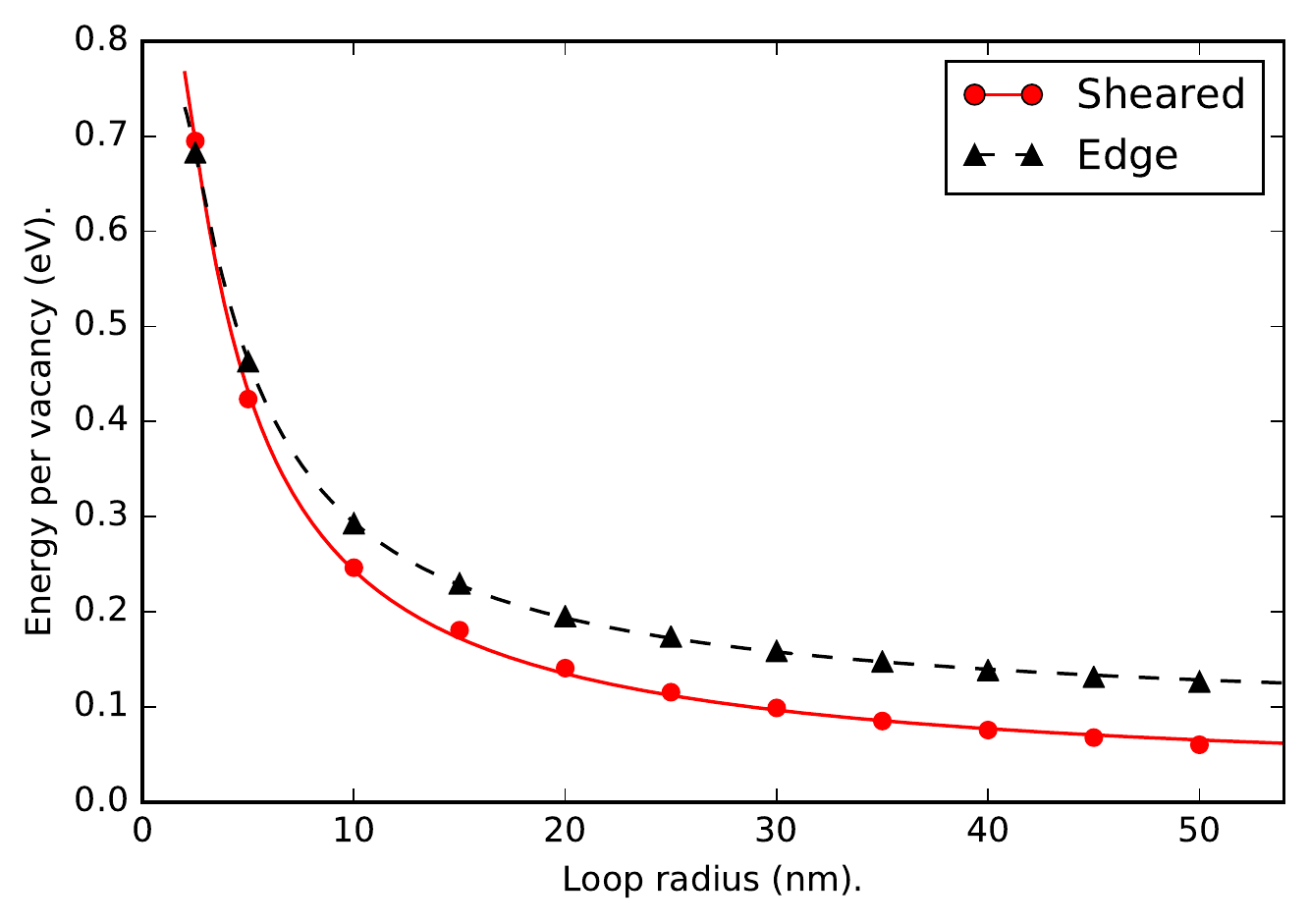}%
\captionsetup{justification=centering,singlelinecheck=false}  %
\caption{}
\end{subfigure}
\begin{subfigure}[b]{0.8\figwidth}
\includegraphics[width=0.7\figwidth]{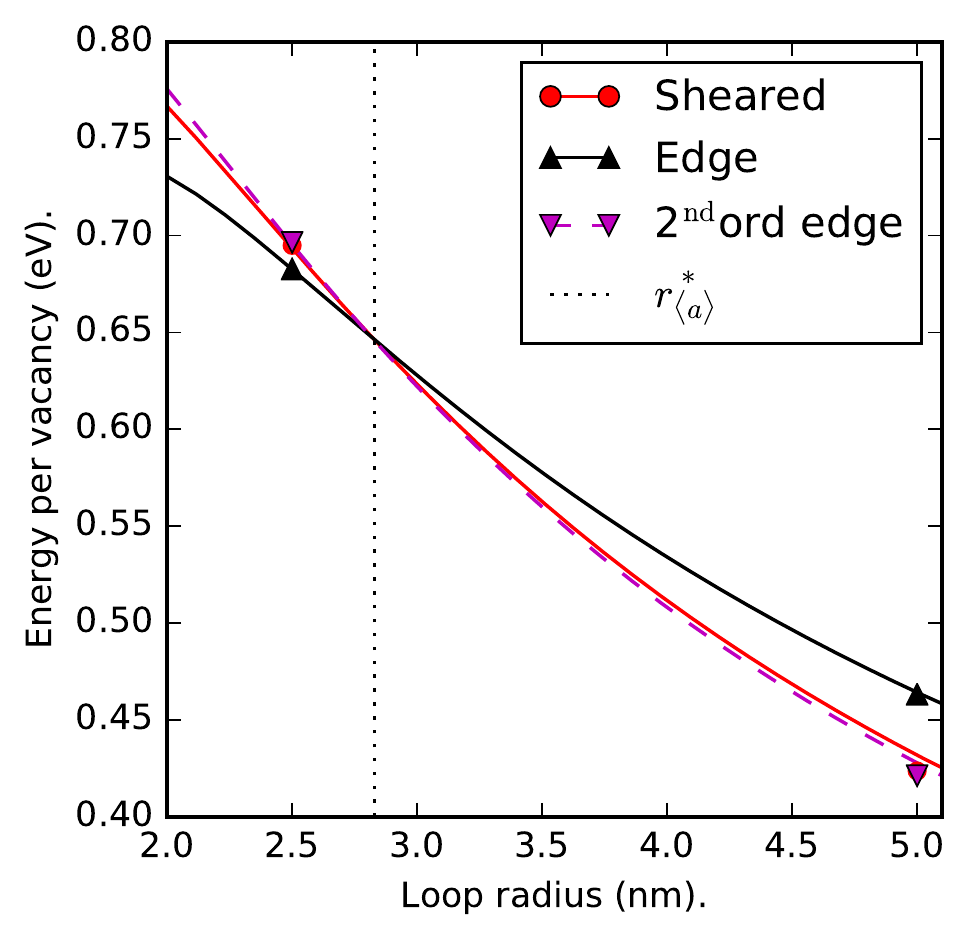}%
\captionsetup{justification=centering,singlelinecheck=false}  %
\caption{}
\label{fig:aLp1prisEdgeVsSheared_b}
\end{subfigure}
\caption{Formation energy per vacancy as a function of radius for edge $\langle a \rangle^{\text{1ord}}_{\text{vac}}$ loops (\bold{b}$=\frac{1}{2} \langle 0 {1} \bar{1} 0 \rangle$) and sheared $\langle a \rangle^{\text{1ord}}_{\text{vac}}$ loops (\bold{b}$=\frac{1}{3} \langle 2 \bar{1} \bar{1} 0 \rangle$).  The unfaulting radius, $r^*_{\langle a \rangle}$, is shown in image (b), which also includes the function for edge $\langle a \rangle^{\text{2ord}}_{\text{vac}}$ loops (\bold{b}$=\frac{1}{3} \langle 2 \bar{1} \bar{1} 0 \rangle$) denoted as `2$^{\text{nd}}$ ord edge' in the legend.}
\label{fig:aLp1prisEdgeVsSheared}
\end{center}
\end{figure}

As a nucleated edge \aloop{} grows by absorbing vacancies, there is an increased thermodynamic driving force for it to unfault via a shearing across the loop plane.  Jostsons et al.~observed sheared \aloops{} \cite{jostsons1977nature}, establishing experimentally that \aloops{} can unfault.  We investigated the radius at which this shear occurs, $r^*_{\langle a \rangle}$, by simulating two series of \aloops{} with various diameters, inhabiting the $(10\bar{1}0)$ plane.  One series had $\textbf{b} = a/2\langle10\bar{1}0\rangle$ and the other had $\textbf{b} = a/3\langle11\bar{2}0\rangle$.  As previously, functions were fitted to the loop energy as a function of radius. Fig.~\ref{fig:aLp1prisEdgeVsSheared}a shows that for all but small radii, sheared \aloops{} have the lowest energy. We calculate the radius $r^*_{\langle a \rangle}$, below which pure edge \aloops{} have the lowest energy (see Fig.~\ref{fig:aLp1prisEdgeVsSheared_b}), to be 2.8~nm, which is close to the $r^*_{\langle a \rangle}$ value determined by Varvenne et al.~of 2.7~nm \cite{varvenne2014vacancy}.  The small values of $r^*_{\langle a \rangle}$ determined by our study and by Varvenne et al. explain why $\langle a \rangle$ loops observed in TEM always have $\textbf{b} = 1/3 \langle11\bar{2}0\rangle$, as resolving loops smaller than $r^*_{\langle a \rangle}$ is difficult.
%

%
\iffalse	%
\subsubsection{section 1.4}
We postulated that vacancies agglomerate into clusters, which later collapse to form dislocation loops.  This process is difficult to observe experimentally but molecular dynamics can simulate this.  We created a supercell containing randomly situated vacancies at a defect density of ??.  This supercell was then subjected to a simulated anneal, by raising it's temperature to XXK over a time period of XXns using a timestep of XXps.  It was held at a constant temperature of XXK for XXns and periodic snapshots were taken of the configuration.  The vacancies, which began as randomly dispersed formed a loop-like structure after the anneal.  The snapshots gave an estimation of how quickly the vacancies coalesced together to form a loop.  A selection of these snapshots can be seen in Fig.~\ref{fig:vacAnnealSnapshots}.

%
%
%
%
%
%
\fi
%
%

%

%
\subsection{Ellipticity of $\langle a \rangle$ Loops.}\label{sec:loopEllip}

\noindent Jostsons et al.~examined \noindent \aloops{} with TEM \cite{jostsons1977nature} and saw them to be elliptical.  The ellipticities they measured are shown in Fig.~\ref{fig:ellipLoopsPlot}.  Ellipticity is defined here as $1-a/b$, where $a$ is the minor axis length and $b$ is the major axis length.  The ellipticity of vacancy \aloops{} increases with diameter, up to around 100~nm where it plateaus at $\sim 0.4$ \cite{jostsons1977nature}.  Interstitial \aloops{} differ in that their ellipticity plateaus earlier at $\sim 0.1$, when their diameter is 20~nm and greater \cite{jostsons1977nature}.  An increase in loop ellipticity increases the loop's dislocation line length and should increase its energy.  This increase in line length must be offset by anisotropy in dislocation line energy which may be due to the anisotropy of the crystal and elastic anisotropy.  Combined with this are possible effects from DAD, which would reduce interstitial \aloop{} ellipticity because preferential diffusion of interstitials along basal planes increases the semi-minor axis and excess vacancies diffusing in $\langle0001\rangle$ reduce the semi-major axis \cite{woo1988theory}.  Woo's 1988 DAD theory predicts the opposite result for vacancy \aloops{}, where the ellipticity is increased \cite{woo1988theory}.  Interstitial \aloops{} and vacancy \aloops{} would be equally and moderately elliptical due to line energy anisotropy alone, but DAD combines with this, to increase vacancy \aloop{} ellipticity and reduce interstitial \aloop{} ellipticity.  Thus, Woo's 1988 DAD theory's predictions fit the observed \aloop{} ellipticities \cite{woo1988theory}.\
  
We have investigated the DAD hypothesis by calculating the optimal ellipticities of $\langle a \rangle^{\ttm{1ord}}_{\ttm{vac}}$ loops, $\langle a \rangle^{\ttm{2ord}}_{\ttm{vac}}$ loops, $\langle a \rangle^{\ttm{1ord}}_{\ttm{sia}}$ loops and $\langle a \rangle^{\ttm{2ord}}_{\ttm{sia}}$ loops.  For each loop type we created, a set of loops with various diameters, and with various ellipticities at each diameter.  For a given diameter, the formation energies, $E_{\textrm{f}}$, as a function of ellipticity ($e$) were fitted to a function:

\begin{equation}\label{eq:elipFunc}
E_{\textrm{f}}(e) = 4 \lambda_\ttm{c} \sqrt{\frac{A}{\pi}} \bigg(\frac{1}{\sqrt{1-e}}  + \alpha \sqrt{1-e}\bigg).
\end{equation}

\noindent Here, $\lambda_\ttm{c}$ and $\lambda_\ttm{a}$ are the line energies per unit length for lines parallel to $[0001]$ and $[2\bar{1}\bar{1}0]$ respectively. $A$ is the area of the loop and $\alpha$ is the ratio $\lambda_\ttm{a}/\lambda_\ttm{c}$.  The minimum of this fitted function gives the optimum ellipticity for that loop type and diameter, as shown in Fig.~\ref{fig:ellipFitting} for a sample diameter.  The optimum ellipticity as a function of loop diameter is shown in Fig.~\ref{fig:ellips}.  Values of $\lambda_{\ttm{c}}$, $\lambda_{\ttm{a}}$ and the average line energy $\bar{\lambda}$, for a 100~nm diameter loop are tabulated in Table~\ref{aLpEtableFitted}.  Here $\bar{\lambda}$ is defined as $(a\lambda_{\ttm{a}} + b\lambda_{\ttm{c}})/(a + b)$.\

\begin{figure}[h!]\begin{center}	%
{\includegraphics[width=\figwidth]{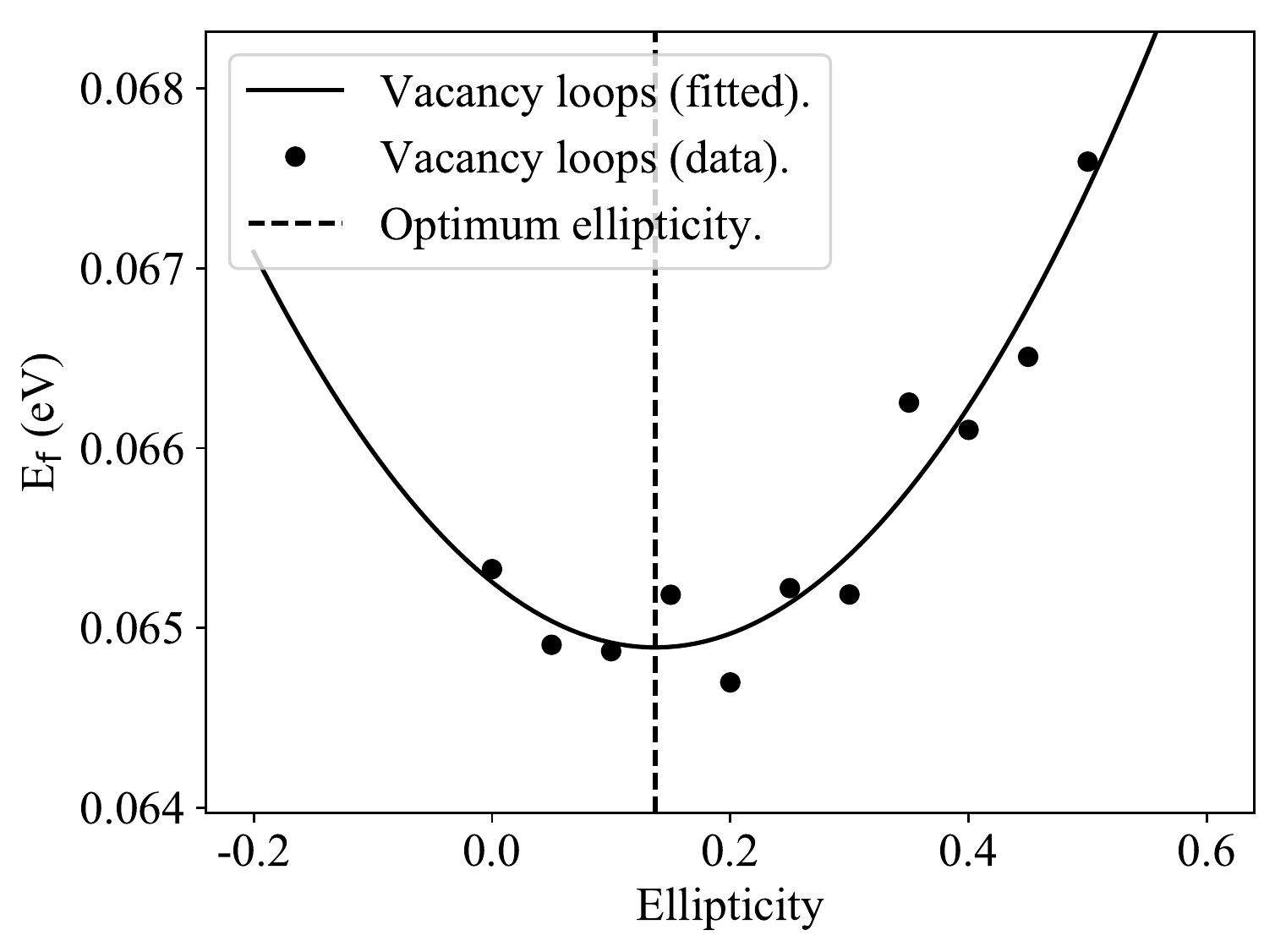}}%
\caption{
Variation of loop energy with ellipticity for the example of 100~nm diameter vacancy $\langle a \rangle$ loops, inhabiting the 1st order prismatic plane.  The solid line shows the fitted function Eq.~\ref{eq:elipFunc}.}
\label{fig:ellipFitting}
\end{center}
\end{figure}

\begin{figure}[htbp!]\begin{center}

{\includegraphics[width=\figwidth]{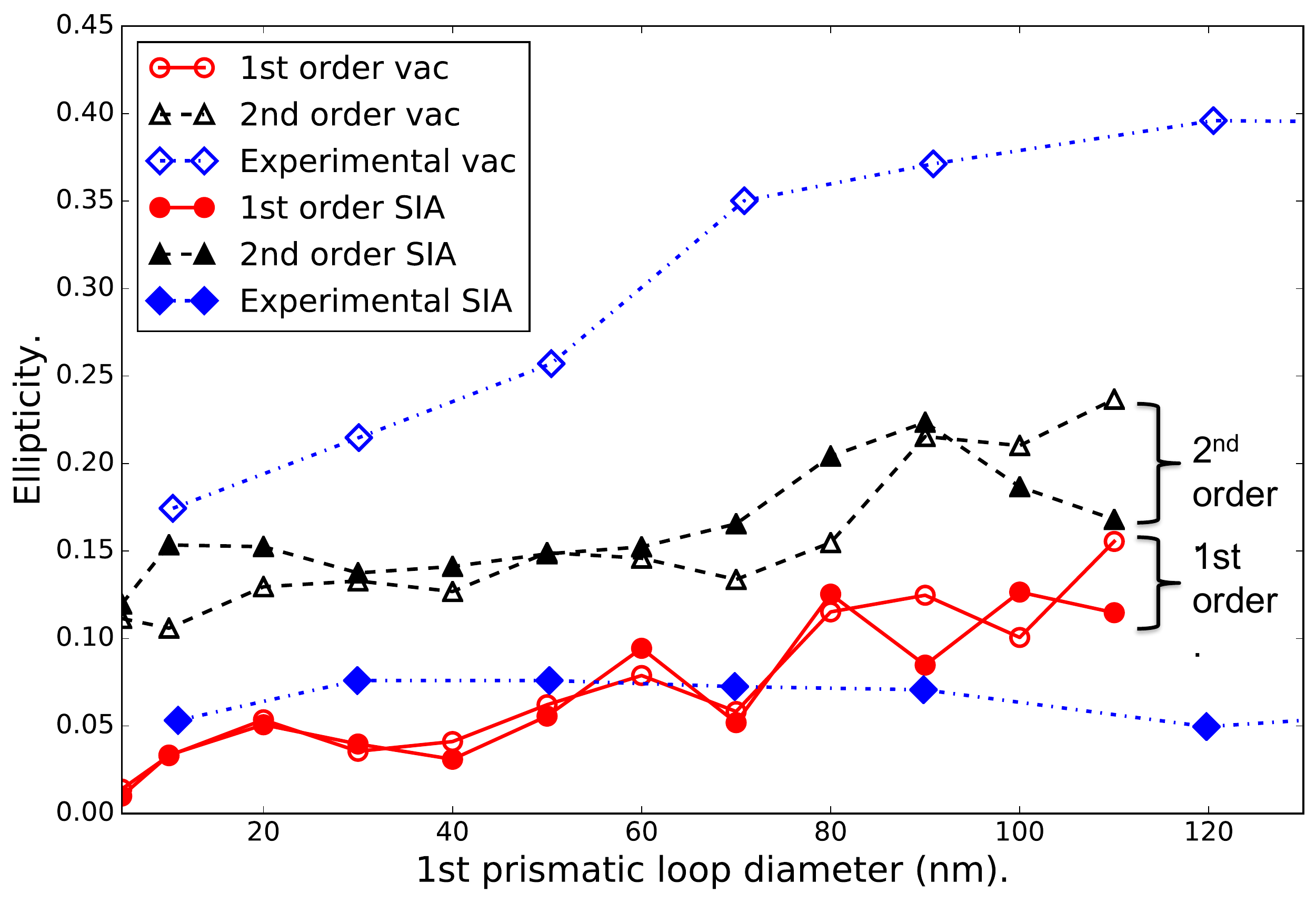}}%
\caption{Ellipticities of experimentally observed and simulated \aloops{}.  The experimental values were taken from Jostsons et al.'s 1977 TEM study \cite{jostsons1977nature}.  Note that ellipticity for 2nd prismatic \aloops{} is expressed as a function of equivalent 1st prismatic loop diameter, where this equivalence is given by defect number.  This provides a fair comparison as the actual 2nd prismatic \aloop{} diameters are smaller than the 1st prismatic diameters by a factor of $\sqrt{3}/2$ for the same number of point defects.}
\label{fig:ellips}
\end{center}
\end{figure}

\begin{table}[h!]
\begin{center}
\begin{tabular}{|c|c|c|c|c|c|}
\hline
 &     &   &  &    \\ [-1.5ex]
Defect  & $\lambda_{\ttm{c}}~(\ttm{eV/nm})$ & $\lambda_{\ttm{a}}~(\ttm{eV/nm})$ & $\bar{\lambda}~(\ttm{eV/nm})$ &  $\alpha$  \\ [0.8ex] \cline{3-5}%
\hline
$\langle a \rangle^\textrm{1ord}_\textrm{vac}$   & 14.3      & 15.9      & 15.3  & 1.11   \\ 
$\langle a \rangle^\textrm{2ord}_\textrm{vac}$   & 13.2      & 16.7       & 14.8  & 1.27   \\ 
$\langle a \rangle^\textrm{1ord}_\textrm{sia}$	 & 14.0	  & 16.0  & 14.9 & 1.15	\\ 
$\langle a \rangle^\textrm{2ord}_\textrm{sia}$	 & 13.3	& 16.3 & 14.4 & 1.23 	\\ [1ex]
    \hline 
\end{tabular}
\caption{The results of fitting Eq.~\ref{eq:elipFunc} for the energies of \aloops{} energetics results. The example of a 100~nm diameter loop is used.  $\alpha$ is the ratio $\lambda_{\ttm{a}}$ : $\lambda_{\ttm{c}}$.}%
\label{aLpEtableFitted}
\end{center}
\end{table}

The results for ellipticity show that for $\langle a \rangle^{\ttm{1ord}}_{\ttm{vac}}$ loops the ellipticity is $\sim$0 for the smallest 5 nm loop.  As $\langle a \rangle^{\ttm{1ord}}_{\ttm{vac}}$ loop diameter increases, ellipticity increases to $\sim0.1$ at 110~nm.  This increase is roughly linear.  The ellipticities for $\langle a \rangle^{\ttm{1ord}}_{\ttm{sia}}$ loops follow a similar pattern to $\langle a \rangle^{\ttm{1ord}}_{\ttm{vac}}$ loops, demonstrating that ellipticity is not a function of loop character.
The results for the 2nd prismatic plane follow a similar trend, although the ellipticities are higher.  For $\langle a \rangle^{\ttm{2ord}}_{\ttm{vac}}$ loops the ellipticity is $\sim0.1$ for the 5 nm loop, increasing linearly to $\sim0.2$ for the 110 nm loop.  As with the 1st prismatic \aloops{}, ellipticity does not change with character.\

The ellipticity fitting data in Table~\ref{aLpEtableFitted} shows that in all cases $\lambda_{\ttm{a}}$ is greater than $\lambda_{\ttm{c}}$.  The values of $\bar{\lambda}$ compare well with those calculated from the energy fitting of \aloops{}, shown in Table~\ref{aLpEtable}.  The differences in these values are partly due to the energy fitting series containing a variety of loop sizes, whereas the values for the ellipticity fitting were taken from a loop with 100~nm diameter. 
  
For vacancy \aloops{}, throughout the experimental data series the ellipticities are greater than those of the simulation series.  This agrees with Woo's 1988 theory which postulates that for vacancy \aloops{} ellipticity will be increased by DAD effects, above that expected due to crystal anisotropy and elastic anisotropy\cite{woo1988theory}.  For interstitial \aloops{}, there is again an agreement with Woo's 1988 theory, as the experimental data shows an ellipticity plateau at $\sim0.1$, which is lower than the ellipticity of the simulated interstitial loops, which do not include the effects of DAD.  Note that for small diameters the $\langle a \rangle^{\ttm{1ord}}_{\ttm{sia}}$ loops have lower ellipticities than for the experimental interstitial data.  However, the experimental loops are a mix of those on the 1st and 2nd prismatic planes.  If the simulated $\langle a \rangle^{\ttm{1ord}}_{\ttm{sia}}$ loops and $\langle a \rangle^{\ttm{2ord}}_{\ttm{sia}}$ loops were similarly mixed, by averaging their ellipticities, the simulated interstitial ellipticities would certainly be higher than the experimental.  The fact that the simulated ellipticities and the experimental ellipticities are closer at small diameters could be an indication that DAD is more influential for larger loops, as these capture more diffusing defects.\

When stating that, considering only elastic anisotropy, vacancy and interstitial \aloops{} would have equal ellipticity, Woo referenced a 1971 TEM study by Brimhall et al.~as justification  \cite{woo1988theory,{brimhall1971microstructural}}.  However, Brimhall et al.'s study was on $\alpha$-Ti, the micrographs only showed a small number of loops and only one of these was identified as having interstitial character.  Thus Brimhall et al.'s study was of limited statistical validity in assessing relationships of ellipticities to loop character.  Additionally, $\alpha$-Ti is a HCP crystal with a $c$:$a$ ratio less than the perfect value and so DAD should alter loop ellipticity.   Our present work provides independent evidence that vacancy and interstitial \aloops{} have equal ellipticity, when only static anisotropy effects are considered, supporting Woo's 1988 theory.

\iffalse	%
\dc{put line study back in, to replace this paragraph: We conducted a further study into \aloop{} ellipticity, using infinite dislocation lines of differing separation \cite{hulse2019simulating}.  Concurring with our work shown here, this showed that $\lambda_{\ttm{a}}$ was greater than $\lambda_{\ttm{c}}$.  This provided additional evidence that ellipticity does not depend on character and that without the influence of DAD, ellipticity depends on the relative formation energy of dislocation line segments in different directions.  Additionally, $\alpha$ increased with separation of the dipole and this provides an explanation as to why \aloop{} ellipticity increases with diameter.}
\fi

\subsubsection{Dislocation dipole energies}\label{sec:InfdislocLineDiffSep}	
\noindent As shown above, the ellipticity of \aloops{} varies with loop diameter.  This may be due to strain interaction from opposing loop segments, making the energy cost of bringing those segments closer together prohibitively expensive.  As diameter increases these opposing segments separate, lowering their mutual strain interaction, to the point where the segments on the minor axis can come together to the equilibrium distance dictated by the relative values of $\lambda_{\ttm{c}}$ and $\lambda_{\ttm{a}}$.  Hence, this process may give an insight not only into ellipticity but also into the effective range of the interaction between dislocation line strain fields.  To probe this, we created a series of infinite $[0001]$ dislocation line dipoles in a thin slab perpendicular to the dislocation line direction and with full periodic boundary conditions. We then varied the dipole separation from 5~nm to 100~nm.  A second series was created, with a dipole line direction of $[2\bar{1}\bar{1}0]$.  The line segments had $\bold{b}=1/3[\bar{1}2\bar{1}0]$ and together they compose a rectangular approximation to an $\langle a \rangle^{\ttm{1ord}}_{\ttm{vac}}$ loop.  We repeated this procedure, using interstitial dipoles (i.e. enclosing a ribbon of interstitial defects) to replicate the line segments of $\langle a \rangle^{\ttm{1ord}}_{\ttm{sia}}$ loops.
\begin{figure}[!ht]\begin{center}
\includegraphics[width=0.9\figwidth]{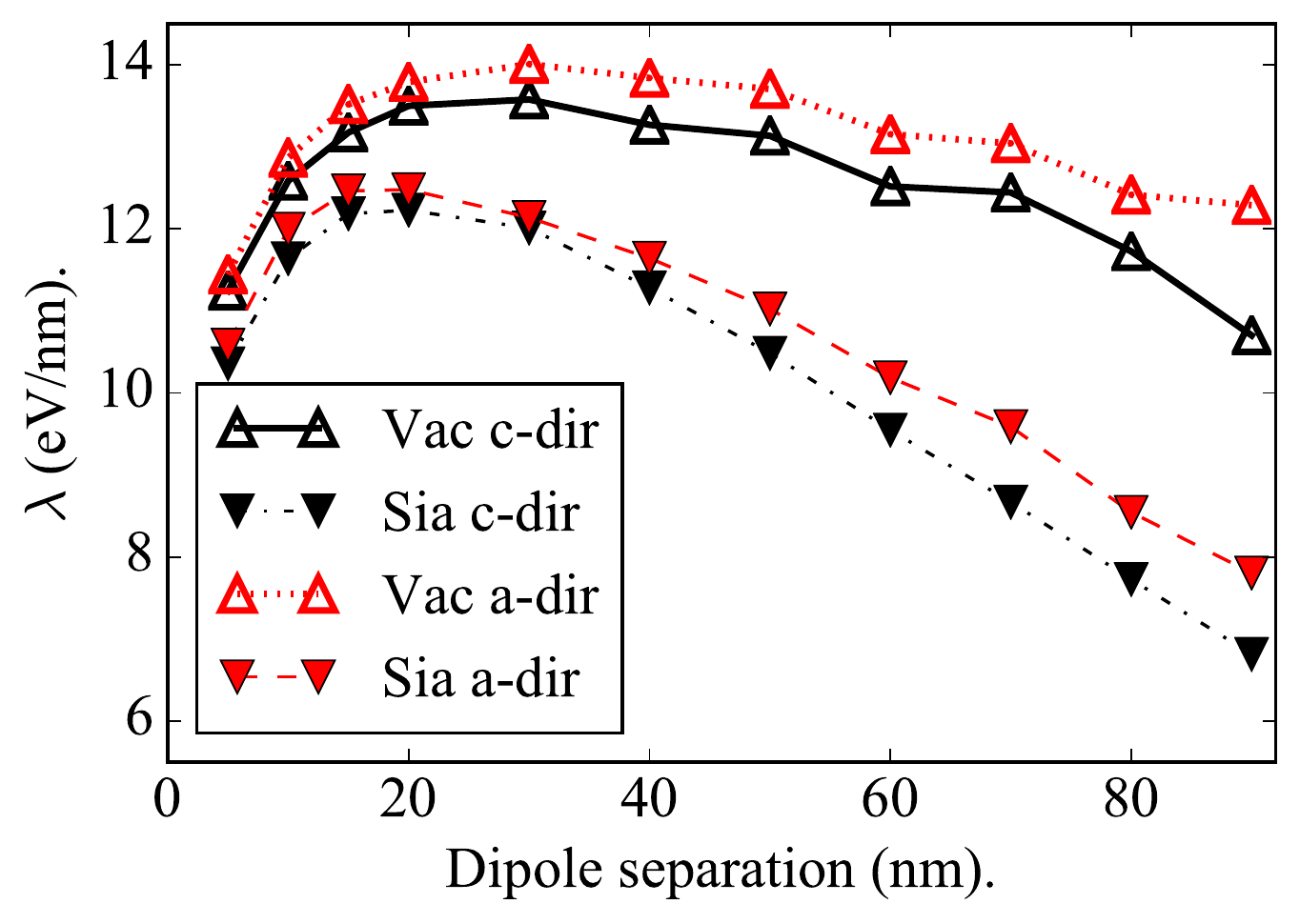} %
\captionsetup{justification=centering,singlelinecheck=false}  %
\caption{Image (a) displays $\lambda$ as a function of dipole separation, for vacancy and interstitial dipoles in the \textit{c}-direction $[0001]$ and \textit{a}-direction $[2\bar{1}\bar{1}0]$.  Image (b) displays ellipticity as a function of dipole separation, for vacancy and interstitial dipoles.}
\label{fig:dipoleE}
\end{center}
\end{figure}
The formation energies per unit length ($\lambda$) of the various dipoles, as a function of separation, are plotted in Fig.~\ref{fig:dipoleE}.  To calculate $\lambda$, we obtained the formation energy of a dipole and subtracted from this the energy of the dipole surface.  The dipole surface is the ribbon of defects that separates the dipoles and it has a formation energy of its own.  Although in theory this surface is unfaulted, in practice it contains a small, finite surface energy.  This surface energy is, as with \aloops{}, due to small disregistry of atoms and so we used the values of $\gamma$ obtained from our study of \aloops{} to calculate the surface energy for the dipoles.\

 For each defect type, $\lambda$ is higher for $[2\bar{1}\bar{1}0]$ than for $[0001]$, as expected.  At small separations, $<$~20~nm, $\lambda$ increases as separation increases.  This is counter to expectations as like dipole lines repel so increasing separation should lower $\lambda$.  We postulate that this occurs because at these small separations the dipole lines' strain fields overlap very strongly.  The resulting strain field is no longer a simple superposition of the two separate dislocations' strain fields, and so $\lambda$ has a lower magnitude than expected.  Above separations $\sim$ 20~nm the dipole's lines act as distinct dislocations and in this regime $\lambda$ decreases as separation increases.  In concurrence with our \aloop{} study in Section~\ref{aLoopPlane}, $\lambda$ is lower for interstitial dipoles than vacancy dipoles.
 
Fig.~\ref{fig:ellipsLoopsAndDipoles} shows the ellipticity as a function of dipole separation implied by the dipole energy calculations.  In this situation the ribbon energy is analogous to the platelet energy and has already been removed, leaving ellipticity dependent on the line energy discrepancy only.  Ergo, we approximate ellipticity as, 1-(1/$\alpha$).
  Ellipticity for vacancy and interstitials is similar, and begins at $\sim$~0.02 for small separations before rising as separation increases.  Ellipticity reaches $\sim$~0.13 at 90~nm for both defect types.  The dipoles are on the 1st order prismatic plane and so we compare these results to the $\langle a \rangle^{\ttm{1ord}}$ loop ellipticity results.  This comparison is displayed in Fig.~\ref{fig:ellipsLoopsAndDipoles} and shows a clear similarity between the ellipticities from our \aloop{} and dislocation dipole studies.  This provides additional evidence that ellipticity does not depend on character and that without the influence of DAD, ellipticity depends on the relative formation energy of dislocation line segments in different directions and will be small.

\begin{figure}[htbp!]\begin{center}
{\includegraphics[width=\figwidth]{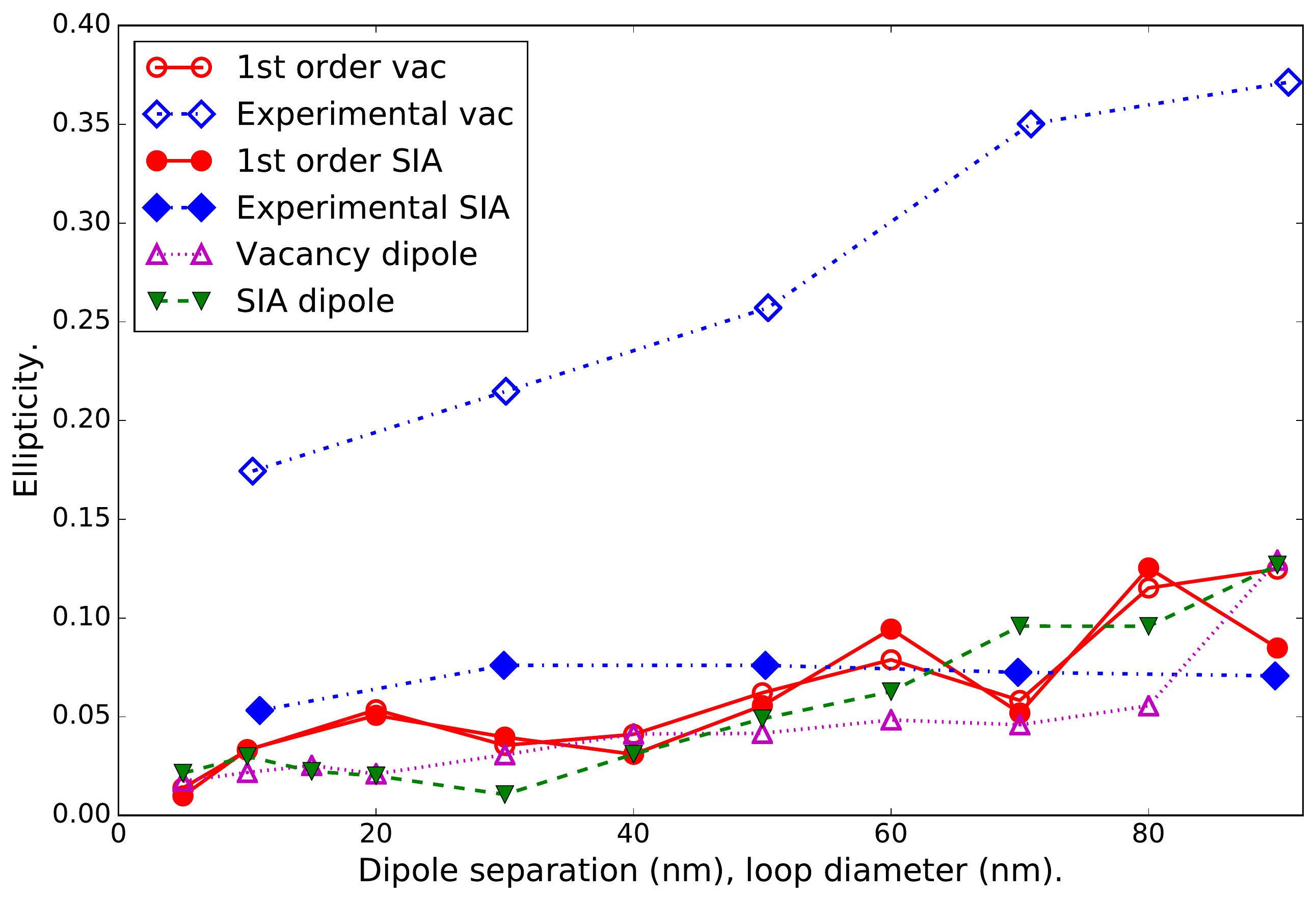}}%
\caption{Ellipticities of dislocation dipoles and $\langle a \rangle^{\ttm{1ord}}$ loops, as a function of separation for the dipoles and loop diameter for the $\langle a \rangle^{\ttm{1ord}}$ loops.  For $\langle a \rangle^{\ttm{1ord}}$ loops, data is included from the simulation results from Section~\ref{aLoopPlane} and experimental results from \cite{jostsons1977nature}.}
\label{fig:ellipsLoopsAndDipoles}
\end{center}
\end{figure}

\subsection{Interstitial $\langle c/2+p \rangle$ loops and $\langle c \rangle$ loops.}\label{sec:SIAcLps_doub_cLps} %

\subsubsection{Interstitial $\langle c/2+p \rangle$ Loops.}
\noindent $\langle c/2+p \rangle$ loops present in irradiated Zr are thought to be vacancy in character \cite{onimus2012radiation,griffiths1987formation}.  This is central to IIG theory that attributes growth to the effective transfer of atoms from basal to prismatic planes \cite{holt1988mechanisms}.  Holt proposed that the loss of atoms from basal planes occurs because \cploops{} form on the basal planes and those \cploops{} are vacancy in character \cite{holt1988mechanisms}.  However, little has been done to determine whether interstitial \cploops{} are feasible.  To probe the stability of interstitial \cploops{} we constructed a series of such loops of various diameters and fitted their formation energies to the same function used for vacancy \cploops{}.  The results and fitted functions are shown in Fig.~\ref{fig:SIAcLps}.  These values can be compared with the formation energy per point defect for vacancy \cploops{}, to give an indication of whether interstitial \cploops{} are realistically stable.  The $\gamma$ and $\lambda$ values pertaining to interstitial \cploops{} and are reported in Table~\ref{sect1.1table}.\

\begin{figure}[htbp!]\begin{center}
{\includegraphics[width=\figwidth]{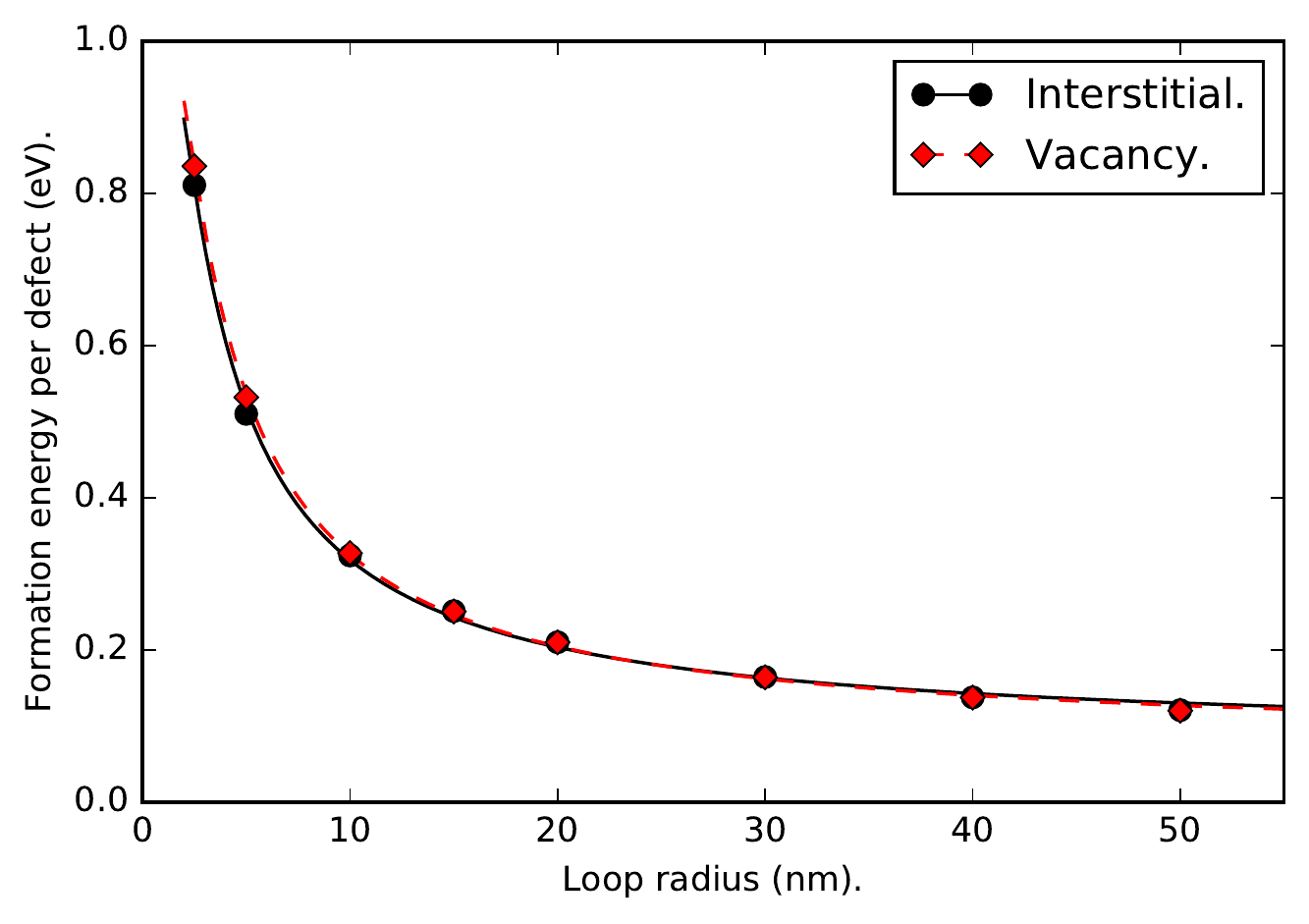}} %
\caption{Formation energy per defect, as a function of radius, for interstitial and vacancy $\langle c/2+p \rangle$ loops.}
\label{fig:SIAcLps}
\end{center}
\end{figure}

Fig.~\ref{fig:SIAcLps} shows that the energies of vacancy $\langle c/2+p \rangle$ loops and interstitial $\langle c/2+p \rangle$ loops are similar, suggesting that interstitial $\langle c/2+p \rangle$ loops are not energetically infeasible.  Thus, their absence may be due to the effects of DAD, which inhibits capture of interstitials by basal planar defects because interstitials diffuse preferentially along basal planes where they are more likely to be captured by prismatic planar defects (i.e. \aloops{}) \cite{woo1988theory}.  However, because little experimental effort has been made to identify the presence of interstitial $\langle c/2+p \rangle$ loops, future experimental work in this area would be valuable, possibly via X-ray diffraction line profile analysis \cite{topping2018investigating}.

\subsubsection{Transition from vacancy \cploops{} to \cloops{}.} %

\noindent As \cploops{} contain an $I_1$ stacking fault there is a contribution to their formation energy that scales as the square of their radius.  However, if two basal layer platelets, rather than one, are removed from the crystal, an unfaulted dislocation loop is created.  As this double layered c-component loop has $\textbf{b}=[0001]$, we term this a \cloop{}.  The dislocation line energy per unit length for a \cloop{}, $\lambda_{\langle c \rangle}$, is expected to be much greater than that of a \cploop{}, $\lambda_{\langle c/2+p \rangle}$, because the burgers vector magnitude of the former is 62\% greater than that of the latter.  Contrasting with this is the SFE contribution, which for the \cploop{} is $\gamma_{I1}$ but for \cloops{} is expected to be close to zero.  Thus, there will there be a critical radius, $r^*_{\langle c \rangle}$, at which a \cloop{} becomes energetically optimal over a \cploop{}.  To find the value of $r^*_{\langle c \rangle}$ we constructed a series of \cloops{} from 50~nm to 150~nm in a simulation box of constant size.  Their energies were fitted, as in Section~\ref{cLpTrans}, using Eq~\ref{eq:loopE}.  This enabled us to find the crossover point of the energy functions for $\langle c/2+p \rangle$ and \cloops{}, which are plotted in Fig.~\ref{fig:doub_cLps}.  As the radius of a \cploop{} is $\sqrt{2}$ bigger than a \cloop{} containing an equivalent number of defects, the crossover was expressed in terms of defect number and occurs at around 100,000 vacancies, or a \cploop{} radius of 56.1~nm.  This crossover radius, whilst large, is not beyond the limit of \cploops{}, which have been observed in TEM studies of Zr as having larger radii than this \cite{topping2018effect,harte2017effect}.  However, Griffiths identified a \cloop{} in Zr via TEM \cite{griffiths1987formation}, suggesting that \cloop{} formation is not impossible.  Griffiths stated that the mechanism of formation for \cloops{} is that, ``vacancies condense on the existing faulted loops forming a double layer and an unfaulted lattice" \cite{griffiths1987formation}.  It could be that this mechanism does not allow the direct transition from a \cploop{} to a \cloop{} and this transformation only occurs when a large \cploop{} is in a region rich in vacancies.  We postulate that an alternative mechanism could occur here, where two \cploops{} coalesce to form a \cloop{}.  The \cloop{} formation phenomenon is worthy of further investigation.

\begin{figure}[!ht]\begin{center}
\includegraphics[width=\figwidth]{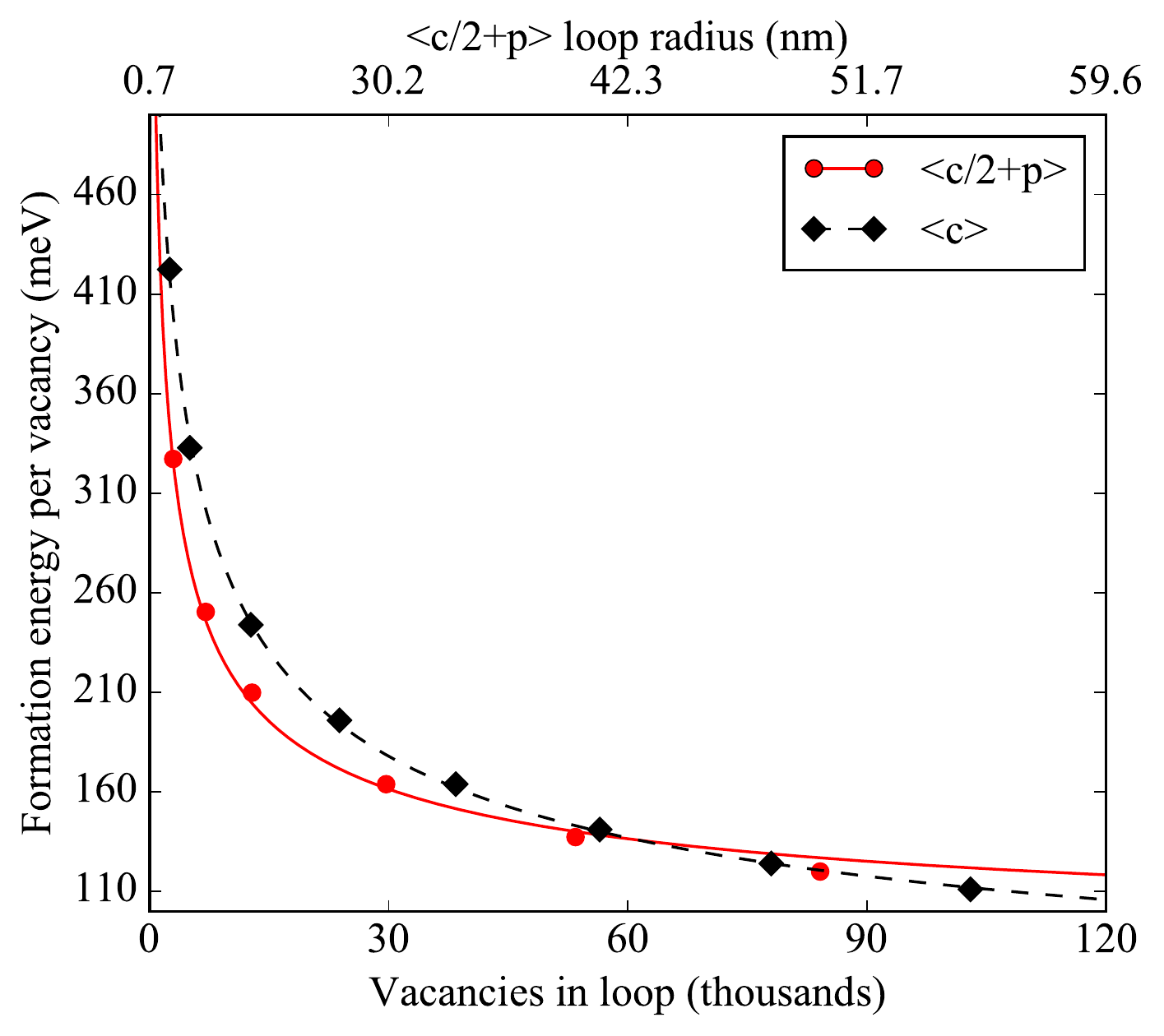} %
\captionsetup{justification=centering,singlelinecheck=false}  %
\caption{Formation energy per vacancy for $\langle c \rangle$ loops, as a function of vacancies in the loop.  The data for $\langle c/2+p \rangle$ loops are shown for comparison, as these may be the defect from which $\langle c \rangle$ loops nucleate \cite{griffiths1987formation}.  The crossover point where the energy per defect of $\langle c \rangle$ loops becomes lower than that of $\langle c/2+p \rangle$ loops is denoted as $r^*_{\ttm{unfaulted}}$.  Data points represent the simulation results and lines are the model fit.}
\label{fig:doub_cLps}
\end{center}
\end{figure}

\subsection{Loop strain fields}\label{sec:strainInter} %

\noindent As briefly explained in Section \ref{sec:loopEllip}, a part of the formation energy of a dislocation loop is due to the interaction of the strain fields of opposing line segments across the centre of the loop. We can consider these opposing segments of the loop as dislocation line dipoles, where the strain interaction is that of one half of the dipole with the strain field of its counterpart.

To study strain fields inside loops, we took the series of loops of varying diameters that were constructed and relaxed in Section \ref{sec:nascLoops}.  From these loops we calculated the elastic strain in the loop centre, using the Ovito software package \cite{stukowski2009visualization}.  

Fig.~\ref{fig:innerStrain} shows the $[0001]$ normal elastic strain values, $\epsilon_{zz}$, at the \cploop{} centre, as a function of loop diameter. These results show that at small diameters, at and below 3~nm, the strain is high, but above 3~nm the strain rapidly decreases.  This threshold, shown as a dashed line in Fig.~\ref{fig:innerStrain}, exists because at small diameters the defect becomes a cluster, containing a highly strained region.  Above the threshold, the defect is a dislocation loop and as its diameter increases the highly strained regions close to the dislocation line move further away from the loop centre. 

\begin{figure}[htbp!]\begin{center}
\includegraphics[width=0.8\figwidth]{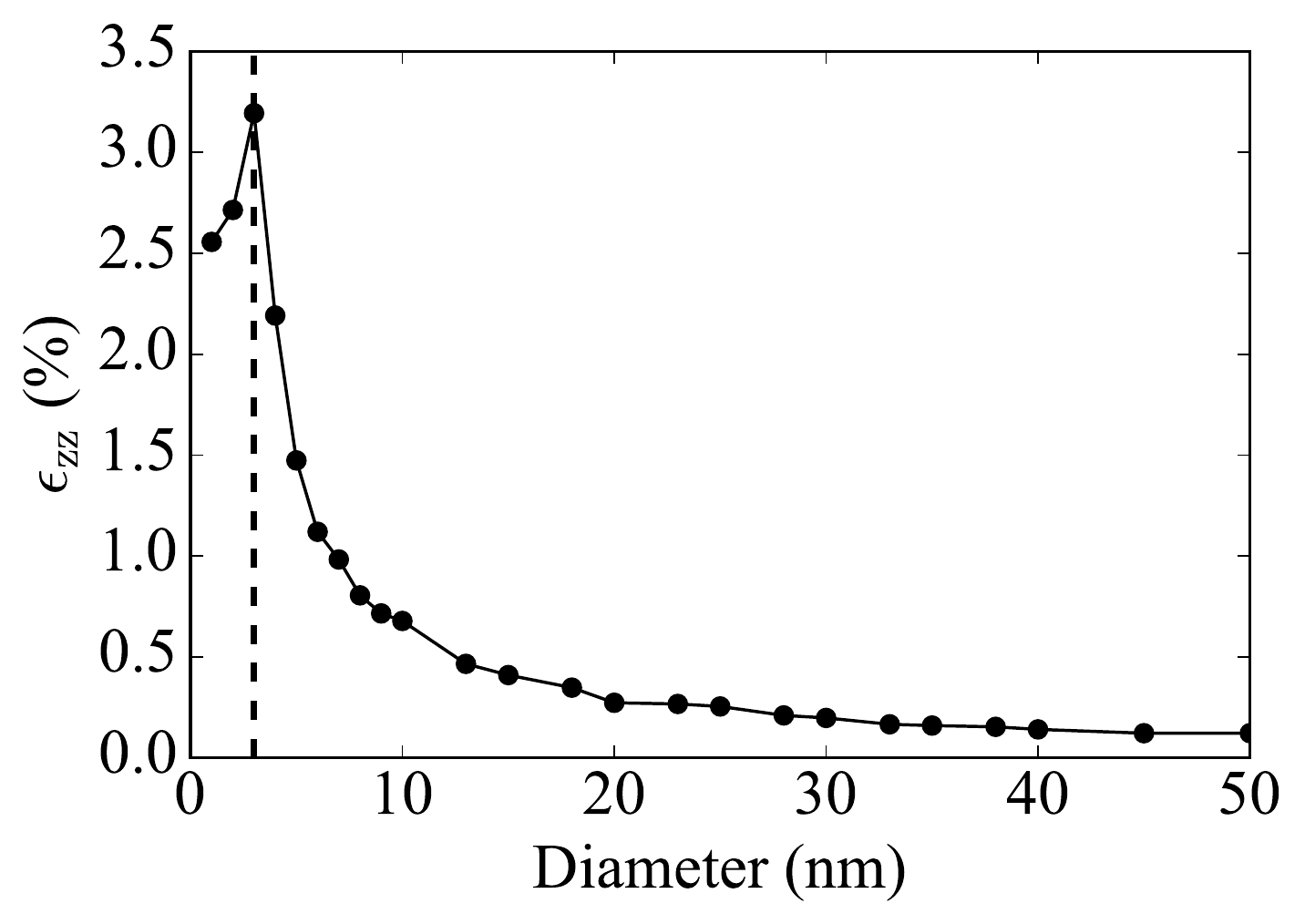}%
\captionsetup{justification=centering,singlelinecheck=false}  %
\caption{Internal $\epsilon_{zz}$ values, in the centre of $\langle c/2+p \rangle$ loops as a function of diameter.  The dashed line shows the threshold, above which the defects behave as loops.  At and below this threshold the defects behave as clusters.}
\label{fig:innerStrain}
\end{center}
\end{figure}

Figure~\ref{fig:constraint} shows maps of the $\epsilon_{zz}$ for \cploops{} of varying sizes. There is a clear trend from the case of small loops (Fig.~\ref{fig:constraint10}), in which the strain is more contained within the loop, to large loops (Fig.~\ref{fig:constraint30}), where the strain field more closely resembles the superposition of the fields of dislocation line segments on opposing sides of the loop. This perhaps provides an explanation for why, with increasing irradiation dose, the \aloops{} increase in number density and decrease in diameter, even though this entails an increase in the formation energy per point defect. The more confined strain fields of smaller loops will reduce the energy of elastic interaction between loops, allowing them to form closer together along their loop normal directions. The increase in energy of the loops per point defect is offset by a reduced interaction energy penalty.

\begin{figure}[!ht]\begin{center}
\begin{subfigure}[b]{0.49\linewidth} 
\includegraphics[width=\linewidth]{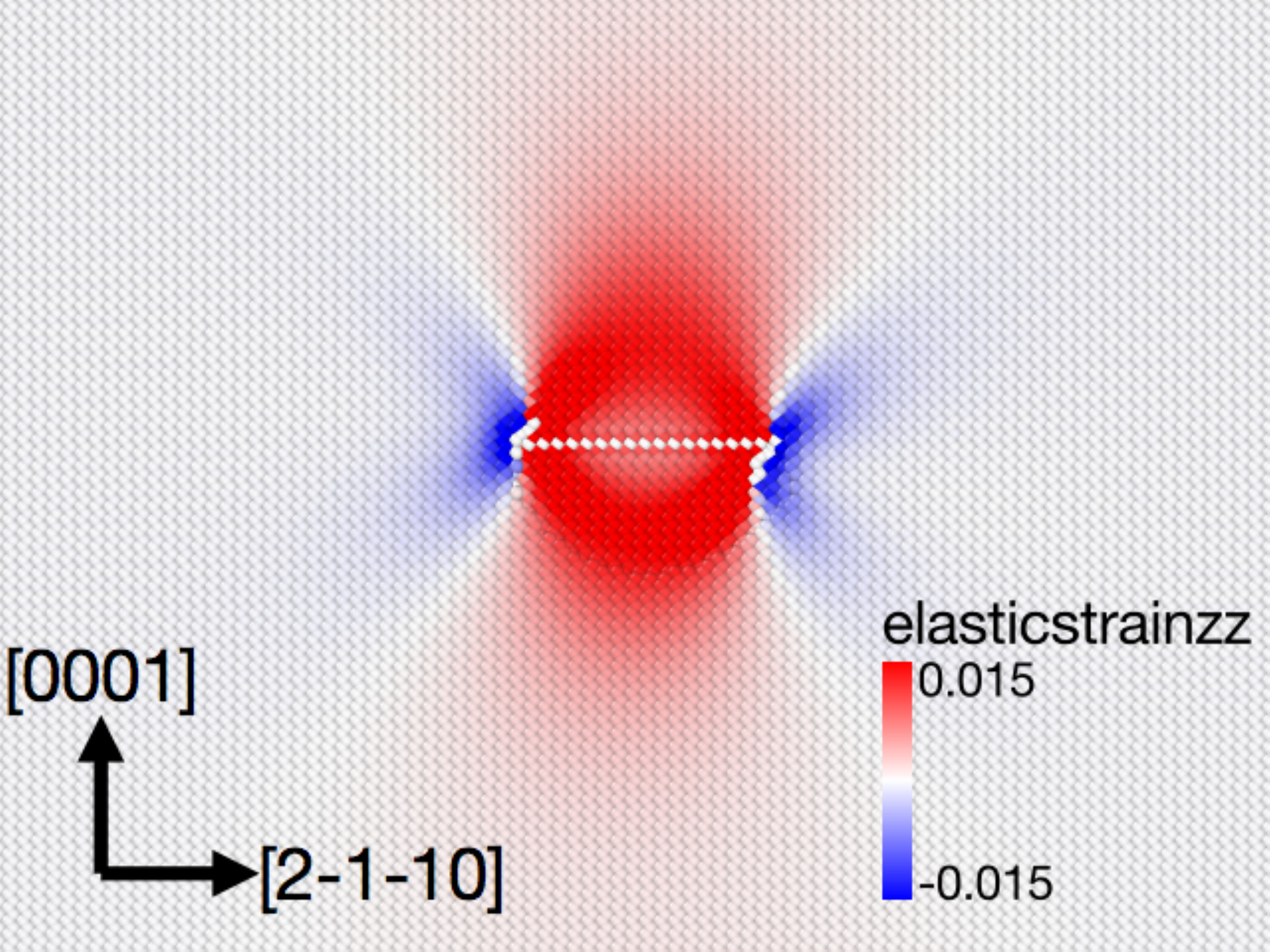}%
\captionsetup{justification=centering,singlelinecheck=false}  %
\caption{10~nm diameter loop.}
\label{fig:constraint10}
\end{subfigure}
\begin{subfigure}[b]{0.49\linewidth} 
\includegraphics[width=\linewidth]{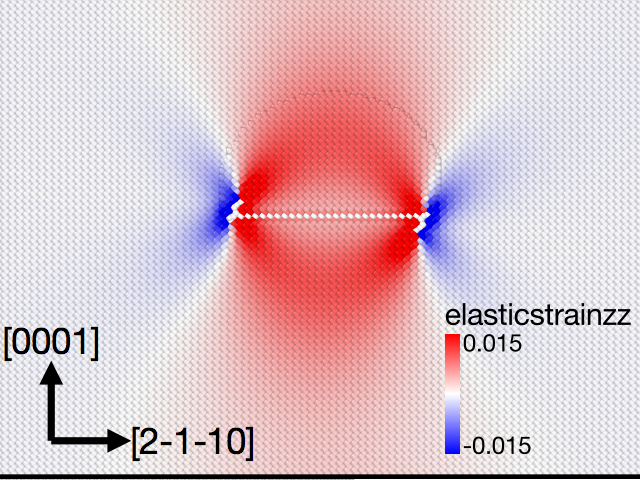}
\captionsetup{justification=centering,singlelinecheck=false}  %
\caption{15~nm diameter loop.}
\label{fig:constraint15}
\end{subfigure}
\
\begin{subfigure}[b]{0.49\linewidth} 
\includegraphics[width=\linewidth]{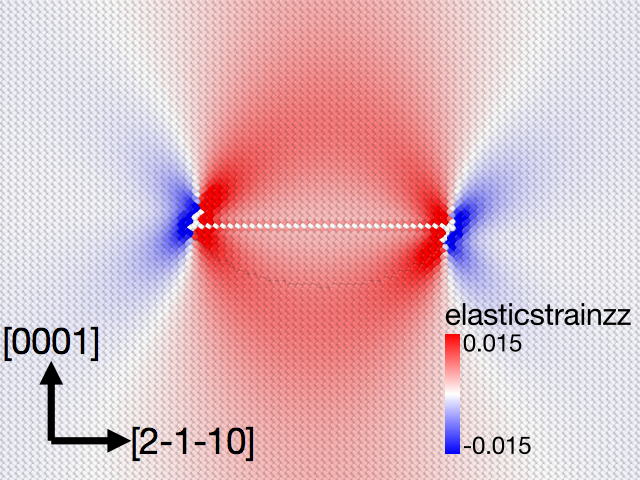}
\captionsetup{justification=centering,singlelinecheck=false}  %
\caption{20~nm diameter loop.}
\label{fig:constraint20}
\end{subfigure}
\begin{subfigure}[b]{0.49\linewidth}
\includegraphics[width=\linewidth]{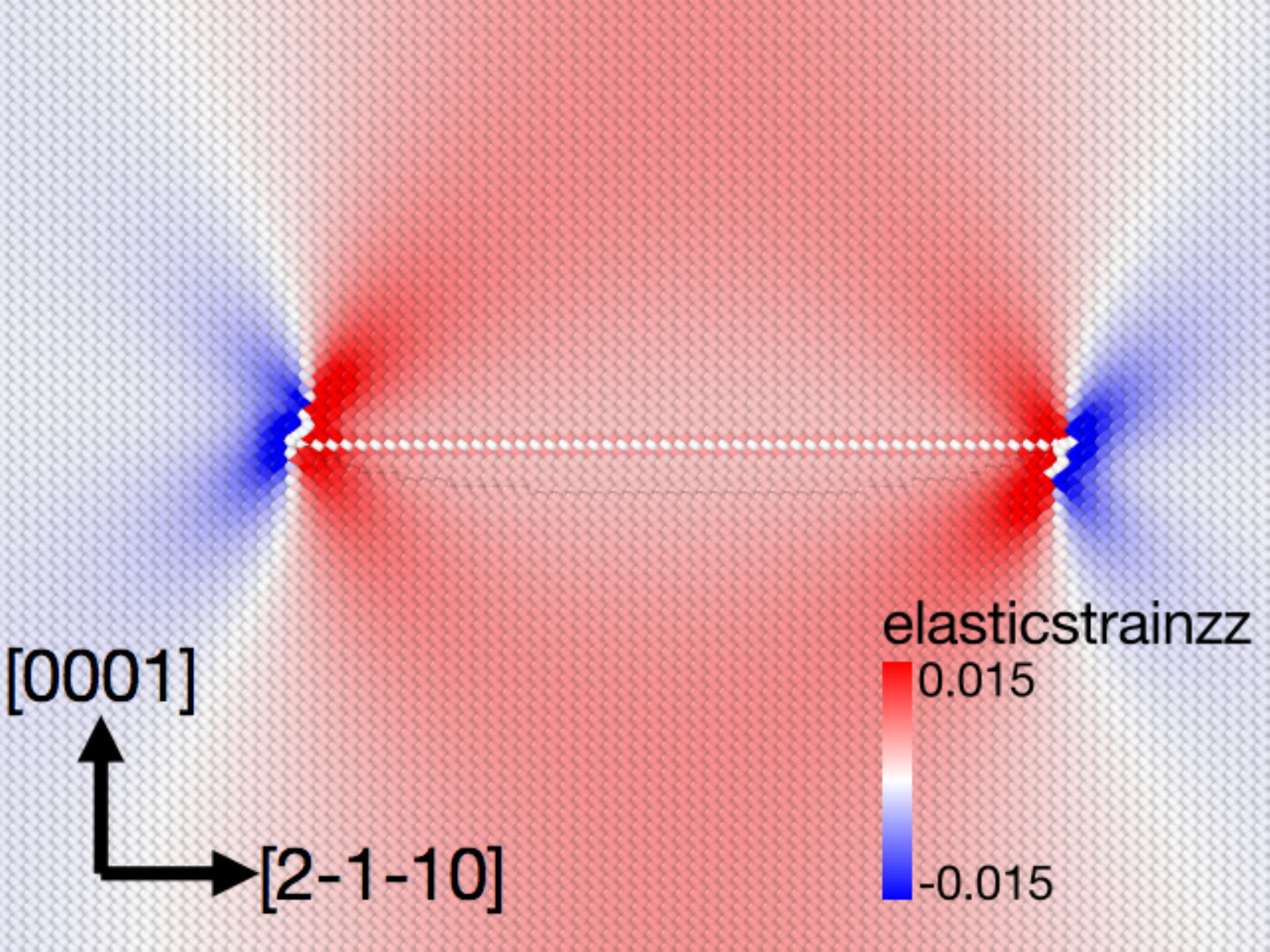}
\captionsetup{justification=centering,singlelinecheck=false}  %
\caption{30~nm diameter loop.}
\label{fig:constraint30}
\end{subfigure}
\caption{Strain maps showing $\epsilon_{zz}$ for $\langle c/2+p \rangle$ loops of various diameters.}
\label{fig:constraint}
\end{center}
\end{figure}

\noindent The rate at which the strain fields decay outside the loops will determine how close together they will tend to form. TEM observations have seen irradiation-induced dislocation loops to be ordered at high densities and typical separation distances have been observed.  For example, Harte et al.\ observed \cploops{} that had a $\sim$50~nm separation along $\langle0001\rangle$ %
To explore this ordering, we constructed a series of \cploops{} in orthogonal supercells, varied their separations along $[0001]$ and calculated the strain at the mid-point between a pair of loops.  The results are displayed in Fig.~\ref{fig:extStraincLps}, in which the strain, $\epsilon_{zz}$, reduces rapidly as the loop separation is increased before plateauing at a very small strain when separations are above $\sim$80~nm.

The experimentally determined loop number densities, line densities and diameters  \cite{barashev2015theoretical,topping2017investigation,topping2018effect,{harte2017effect}} give us an estimation of the spacing between \cploops{} of 40 - 50~nm, and this corresponds to strains between $\epsilon_{zz} = 0.422\%$ and $\epsilon_{zz} = 0.207\%$.  This range is shown by the dotted lines in Fig.~\ref{fig:extStraincLps}.  Clearly there is a complicated interplay when loop separation increases, as elastic interaction energy decreases, but formation energy increases because there is a lower loop density (assuming a fixed number of point defects). 

\begin{figure}[htbp!]\begin{center}
{\includegraphics[width=0.75\figwidth]{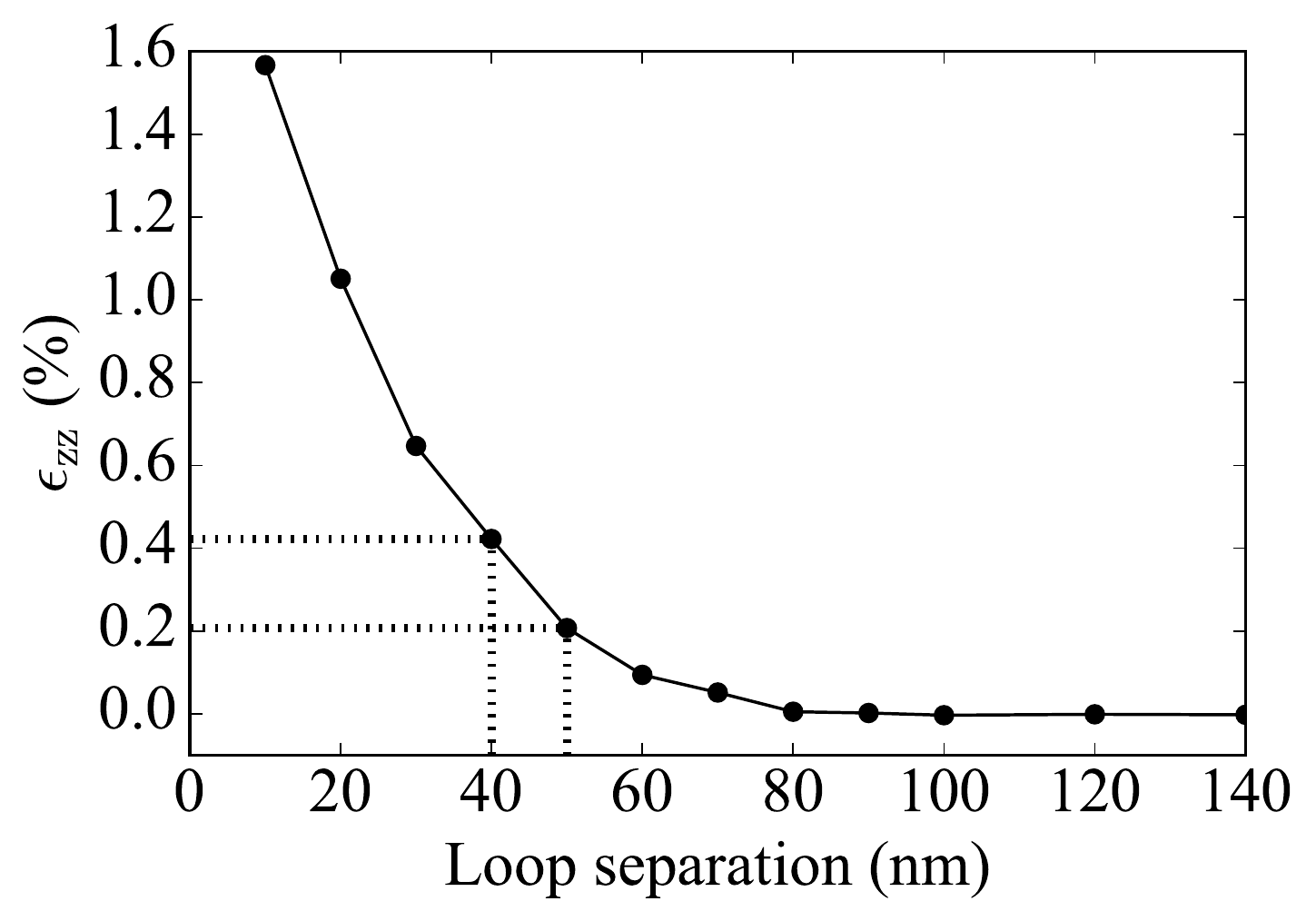}}%
\caption{External $\epsilon_{zz}$ values, in between two $\langle c/2+p \rangle$ loops as a function of separation of the loops.  The dotted lines shows the \cploop{}  separation range that we estimate here as between 40 - 50~nm.}
\label{fig:extStraincLps}
\end{center}
\end{figure}

\section{Conclusions.}
\noindent Increases in computing power now allow routine simulation of dislocation loop populations at experimentally relevant scales and we believe that our simulated dislocation loops are bigger than those previously simulated.  We have been able to study the effects of dislocation lines and loop planar faults in combination, and to include the effects of strain on opposing loop segments.  We believe this study is also the first to simulate elliptical loops.  These techniques have enabled us to study dislocation loop sizes, configurations and densities in detail and this has revealed new insights in the loop behaviour.  The most pertinent of these are:

\begin{itemize}
\item Our calculated value for the critical radius at which a \chloop{} transforms to a \cploop{}, $r^*_{\text{HE}\rightarrow \text{I1}}$, is 3.2~nm and this is close to that reported by Varvenne et al.~\cite{varvenne2014vacancy}.  Thus, the radius at which we expect the $\text{HE}\rightarrow\text{I1}$ transformation to occur is small and is consistent with the notion that the \chloop{} is the precursor to the \cploop{}.

\item The energies of the 1st order sheared and 2nd order edge \aloops{} are very similar, which may be the reason these loops do not inhabit a definite plane, but have a distribution between the two. The results we have presented suggest that \aloops{} form on the 1st prismatic plane as pure edge loops, shear to create 1st order prismatic sheared loops and then rotate to inhabit the 2nd order prismatic plane as pure edge loops.  However, their rotation onto the 2nd order prismatic plane may be inhibited by obstacles and there may be insufficient thermodynamic driving force to overcome these, due to the small difference in formation energies between 1st order sheared \aloops{} and 2nd order edge \aloops{}.

\item We have shown that based on energy considerations alone ellipticity is the same for interstitial and vacancy \aloops{}.  This was stated by Woo, but was not supported by strong evidence \cite{woo1988theory}.  Our results support the predictions of Woo's 1988 theory by showing that without the effects of DAD, vacancy \aloop{} ellipiticity is lower than that experimentally observed and interstitial \aloop{} ellipticity is higher.  The results from our \aloop{} study and our dislocation dipole study agree on this.  Ellipticities for \aloops{} on 2nd order prismatic planes are greater than those on 1st order prismatic planes.

\item We confirmed, in the \aloop{} ellipticity study of Section~\ref{sec:loopEllip}, that the energy per line length is greater in $a$-directions than in $c$-directions for dislocations with $\bold{b}=1/3\langle1\bar{2}10\rangle$.  This creates the correct energetics for ellipses with the major axis parallel to [0001].  Our study of dislocation dipoles revealed that the relative energy differences between these increase with line separation, explaining why \aloops{} become more elliptical as their diameter increases.

\item Interstitial \cploops{} are not energetically infeasible, so their existence cannot be ruled out.  However, DAD effects may prevent interstitial \cploops{} from growing.  This warrants further experimental study.

\item A \cploop{} with radius greater than $\sim$50~nm would lower its energy if it could transform into a \cloop{}.  However, the mechanism for this transformation to occur may be restrictive.  This aspect of $c$-component loops is ripe for further investigation.

\item Our results show that \aloops{} have lower formation energy per defect than \cploops{} of the same size.  This indicates that \cploops{} should not occur, but evidently they do.  We postulate that many smaller \aloops{} transform into a large \cploop{} because the formation energy per defect is lower for a large \cploop{} than for the same defects accommodated in many smaller \aloops{}.  Experimental results of Topping et al.~\cite{topping2018effect} agree with this notion as for Zircaloy-2 irradiated to 4.7dpa, \aloops{} had a diameter of $\sim$10~nm and \cploops{} a diameter of $\sim$100~nm.  Our model predicts, for these diameters, that the energy per vacancy will be lower for \cploops{} than for \aloops{}.

\indent Harte et al.~\cite{harte2017effect} produced Bright Field Scanning Transmission Electron Microscope (BF-STEM) images of proton irradiated Zircaloy-2, shown in Fig.~\ref{fig:BFStemLoops}.  These provide additional evidence for the idea that \cploops{} originate from ordered arrays of \aloops{} as \cploops{} appear anti-correlated to \aloops{} in the \aloops{}' rafts along the basal plane trace, creating gaps in these rafts.

\begin{figure}[htbp!]\begin{center}
{\includegraphics[width=1.0\figwidth]{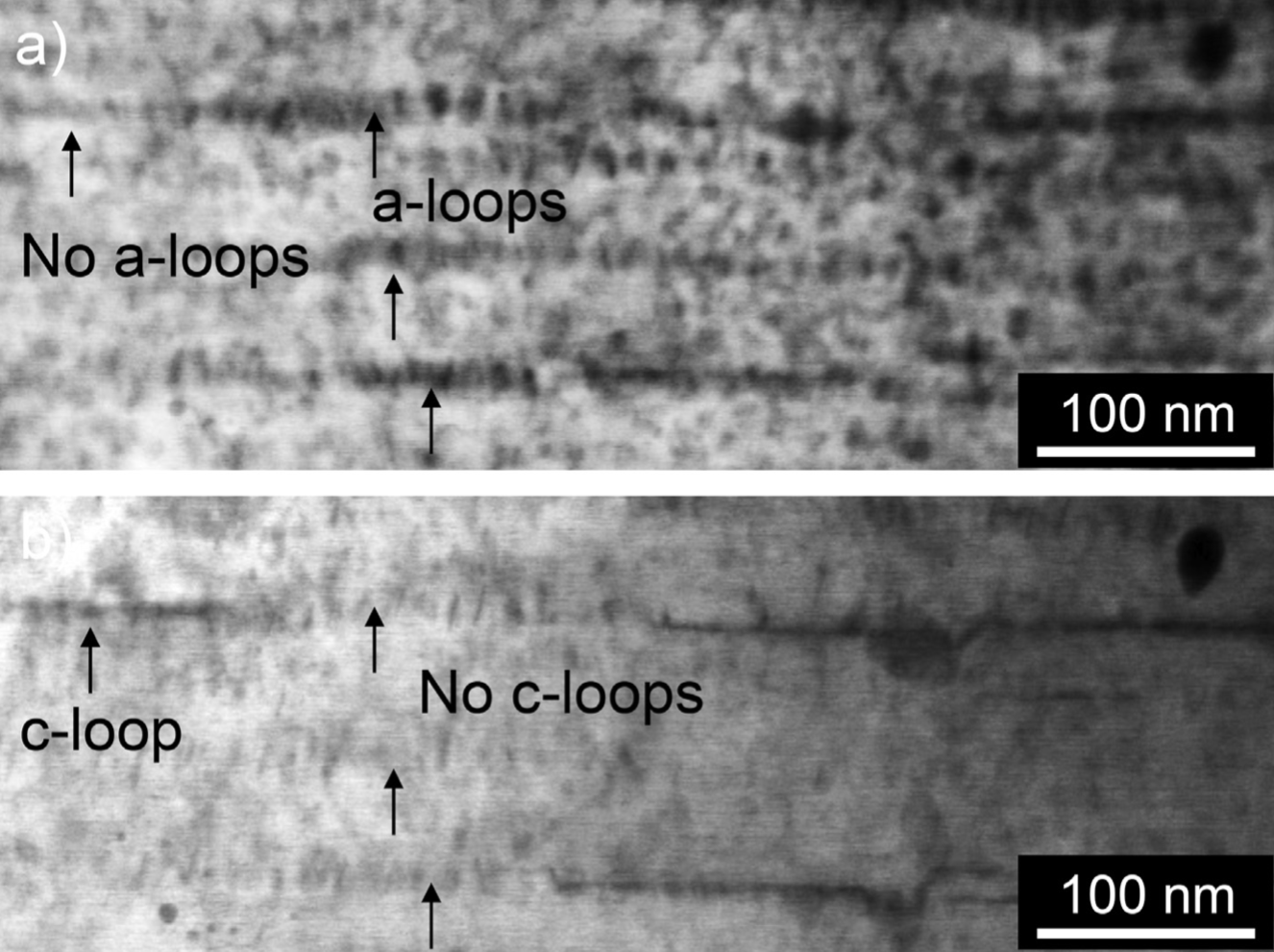}}
\caption{BF-STEM images of proton irradiated Zircaloy-2, where image a) was taken along $\langle 11\bar{2}0 \rangle$ and for b) \bold{g} = 0002.  Therfore, \aloops{} and \cploops{} are visible in a), but \aloops{} are invisible in b) (Reproduced from Harte et al.~\cite{harte2017effect}, available at https://doi.org/10.1016/j.actamat.2017.03.024, under the terms of the Creative Commons Attribution Licence (CC BY)  https://creativecommons.org/licenses/by/4.0/.).}
\label{fig:BFStemLoops}
\end{center}
\end{figure}

\item We have shown that in small loops, the high-strain lobes emanating from the dislocation lines overlap strongly.  As the diameter increases the strain extends further through the crystal and the strain at the loop centre decreases.  However, as \aloop{} diameter decreases the strain field is highly confined close to the loop and does not interact strongly with that of neighbouring loops.  This allows \aloops{} to position themselves closer together, which we postulate is the reason that \aloop{} diameters reduce as irradiation proceeds and more point defects need accommodating in a given volume.

\end{itemize}

\section{Acknowledgements.}
\noindent For providing us with sponsorship and support, we express gratitude to EDF and particularly Antoine Ambard of EDF.  Additionally, we thank the Engineering and Physical Sciences Research Council for providing us with funding through Doctoral Training Centre in Advanced Metallic Systems grant (EP/G036950/1). CPR was funded by a University Research Fellowship of The Royal Society. Calculations made use of the University of Manchester's Computational Shared Facility.

\section{References}
\bibliographystyle{unsrt}
\bibliography{./references/refsPaper1journal}

\appendix
\section{Dislocation Loop Construction}\label{app:dislocationconstruction}
Dislocation loops were created by first removing or adding a platelet of atoms, depending on the character of the loop.  The surrounding atoms were then displaced according to a model displacement field 

\begin{equation}
\textbf{u}_0(\textbf{r}) = \bold{b} ~\alpha(\mu) \beta({d}),
\end{equation}
where \textbf{r} is the initial position of an atom, \bold{b} is the burgers vector of the loop and $\alpha(\mu)$ and $\beta(d)$ are two functions to be defined below. The $\alpha$ function is given by:

\begin{equation}
\alpha(\mu) =\begin{cases}
\scalebox{1.35}{$\frac{\vline \mu \vline}{\mu}$}~0.5 &\text{ for } \mu < S;\\
\scalebox{1.35}{$\frac{\vline \mu \vline}{\mu}~\frac{0.5}{m \mu - c}$} &\text{ for } S < \mu < M;\\
0 & \text{ for } \mu > M.
\end{cases}
\end{equation}
\noindent $\mu = \textbf{r} \cdot \hat{\textbf{n}}$ is the distance from the habit plane of the dislocation loop to an equivalent plane in which the target atom lies as shown in Fig.~\ref{fig:alphaFunction}a, $\hat{\textbf{n}}$ being the unit normal to the loop plane. $S$ is the inter-planar spacing of the loop's habit plane and $M$ = (supercell length / 2) - $S$ sets the effective range of the initial displacement field. The decay in strain given by $\alpha$ has the form $1/\mu$, which is a reasonable approximation for straight line dislocations \cite{landau1959course}.  $c$ is an intercept value, ensuring that $\alpha(\mu)$ is continuous at  $\mu = S$.  To ensure that the initial displacement field decays along the whole of the available simulation box length, a gradient $m$ is included, and this is inversely proportional to the box length.
 
The second function, $\beta(d)$, controls how the displacement varies around the edge of a virtual feature termed the `displaced loop', displayed in Fig.~\ref{fig:alphaFunction}b.  All target atoms in a certain equivalent plane share a displaced loop, which has a shape identical to that of the dislocation loop.  The displaced loop lies on an infinite projection of the burgers vector of the dislocation loop originating at the dislocation loop centre.  For $\beta(d)$, we used a sigmoid function given by

\begin{equation}\label{eq:sigmoid}
\beta(d) = \frac{1}{\exp(\frac{d}{{w}/2}) + 1}.
\end{equation}

\noindent The target atom's distance from the displaced loop edge is $d$, and $w$ is the dislocation core radius.  Dislocation core widths of different crystals vary, but are typically between 1 and 5 lattice parameters \cite{hull2001introduction}. Here we use the value of 1 lattice parameter for the core width.

SIA dislocation loops were created in a similar manner, except in this case the atomic planes were distorted outwards, relative to the loop, then a platelet of atoms was inserted into the void.

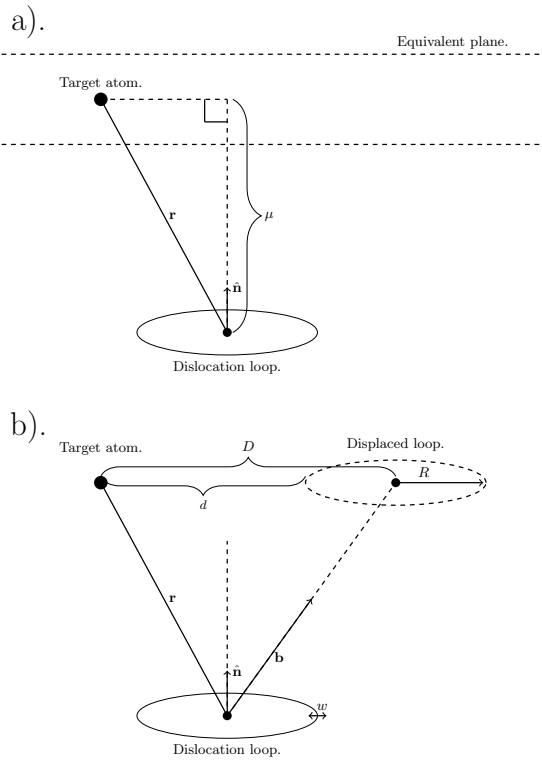
\begin{figure}[htbp!]\begin{center}
\scalebox{0.6}{
\begin{tikzpicture}
\def\c{5.168}
\def\k{5.60155}
\def\xCentre{5}
\def\yCentre{19.5}
\node[above] at (.6,\yCentre+\c+1.25){\huge{a).}};  %
\draw (\xCentre,\yCentre) ellipse (2cm and 0.5cm)node [black,yshift=-22pt] {Dislocation loop.};	  %
\fill[black](\xCentre,\yCentre) circle(0.1);
\draw[dashed,-,thick](\xCentre,\yCentre)--(\xCentre,\yCentre+\c);  %
\draw[->,thick](\xCentre,\yCentre)--(\xCentre,\yCentre+1.);  %
\node[right] at (\xCentre,\yCentre+1.){$\hat{\textbf{n}}$};  %
\fill[black](\xCentre-\k*1./2.,\yCentre+\c) circle(0.15);	%
\node[above] at (\xCentre-\k*1./2.,\yCentre+\c+0.1){Target atom.};  %
\draw[->,thick](\xCentre,\yCentre)--(\xCentre-\k*1./2.,\yCentre+\c);  %
\node[right] at (\xCentre-\k*.5/2.,\yCentre+\c*.5){$\textbf{r}$};  %
\draw [decorate,decoration={brace,amplitude=18pt,mirror},xshift=3.5pt,yshift=0pt](\xCentre,\yCentre) -- (\xCentre,\yCentre+\c)node [black,midway,xshift=23pt] {$\huge{\mu}$};	  %

\draw[dashed,-,thick](0,\yCentre+\c-1.)--(12,\yCentre+\c-1.);  %
\draw[dashed,-,thick](0,\yCentre+\c+1.)--(12,\yCentre+\c+1.);  %
\node[above] at (10,\yCentre+\c+1.){{Equivalent plane.}};  %
\draw[dashed,-,thick](\xCentre-\k*1./2.,\yCentre+\c)--(\xCentre,\yCentre+\c);  %
\draw[-,thick](\xCentre-1./2.,\yCentre+\c)--(\xCentre-.5,\yCentre+\c-.5);  %
\draw[-,thick](\xCentre,\yCentre+\c-.5)--(\xCentre-.5,\yCentre+\c-.5);  %
\def\yCentre{11}
\node[above] at (.6,\yCentre+\c+0.75){\huge{b).}};  %
\draw (\xCentre,\yCentre) ellipse (2cm and 0.5cm)node [black,yshift=-22pt] {Dislocation loop.};	  %
\fill[black](\xCentre,\yCentre) circle(0.1);
\draw[dashed,-,thick](\xCentre,\yCentre)--(\xCentre,\yCentre+\c*.75);  %
\draw[->,thick](\xCentre,\yCentre)--(\xCentre,\yCentre+1.);  %
\node[right] at (\xCentre,\yCentre+1.){$\hat{\textbf{n}}$};  %
\draw[->,thick](\xCentre,\yCentre)--(\xCentre+\k/3.,\yCentre+\c/2.);  %
\node[right] at (\xCentre+.5*\k/3.,\yCentre+.5*\c/2.){$\textbf{b}$};  %
\draw[dashed,-,thick](\xCentre,\yCentre)--(\xCentre+\k*2./3.,\yCentre+\c);  %
\draw[dashed,-,thick] (\xCentre+\k*2./3.,\yCentre+\c) ellipse (2cm and 0.5cm); %
\fill[black](\xCentre+\k*2./3.,\yCentre+\c) circle(0.1);	%
\node[above] at (\xCentre+\k*2./3.,\yCentre+\c+.6){Displaced loop.};  %
\fill[black](\xCentre-\k*1./2.,\yCentre+\c) circle(0.15);	%
\node[above] at (\xCentre-\k*1./2.,\yCentre+\c+0.5){Target atom.};  %
\draw[->,thick](\xCentre,\yCentre)--(\xCentre-\k*1./2.,\yCentre+\c);  %
\node[right] at (\xCentre-\k*.5/2.,\yCentre+\c*.5){$\textbf{r}$};  %
\draw [decorate,decoration={brace,amplitude=12pt,mirror},yshift=4pt](\xCentre-\k*1./2.,\yCentre+\c) -- (\xCentre-2+\k*2./3.,\yCentre+\c)node [black,midway,yshift=-18pt] {\footnotesize$d$};	  %
\draw [decorate,decoration={brace,amplitude=12pt},yshift=4pt](\xCentre-\k*1./2.,\yCentre+\c) -- (\xCentre+\k*2./3.,\yCentre+\c)node [black,midway,yshift=19pt] {$D$};  %
\draw[<->,thick](\xCentre+1.8,\yCentre)--(\xCentre+2.2,\yCentre)node[midway,above,xshift=3pt]{${w}$};  %
\draw[->,thick](\xCentre+\k*2./3.,\yCentre+\c)--(\xCentre+1.94+\k*2./3.,\yCentre+\c)node[midway,above,xshift=-10pt]{${R}$};  %
\end{tikzpicture}
}
\caption{The constructions used in obtaining arguments for the $\alpha$ and $\beta$ functions.  The distance $\mu$, used to determine the strain decay away from the loop, is shown in a).  D is the distance between the target atom and the displaced loop centre.  The displacements around the edges of the displaced loops are controlled by the distance d.}
\label{fig:alphaFunction}
\end{center}
\end{figure}

\end{document}